\documentclass[12pt]{article}
\usepackage{bbding}
\usepackage{amsmath}
\usepackage{amssymb}
\usepackage{amsfonts}
\usepackage{mathrsfs}
\usepackage{mathrsfs, amssymb}
\usepackage[dvips]{graphicx}
\usepackage{booktabs}
\usepackage[perpage,symbol*]{footmisc}

\newcounter{theorem}
\newtheorem{theorem}{Theorem}[section]
\newcounter{lemma}

\newcounter{remark}
\newtheorem{remark}{Remark}[section]
\newcounter{example}

\newcounter{definition}

\newcounter{corollary}
\newtheorem{corollary}{Corollary}[section]
\newcounter{proposition}

\def\cov{\mbox{cov}}

\def\drow{\stackrel{d}{\longrightarrow}}
\def\prow{\stackrel{\textstyle p}{\longrightarrow}}

\begin{document}
\baselineskip 18pt
\begin{center}
{\large\bf  An efficient and doubly robust empirical likelihood approach for estimating equations with missing data}
\end{center}
\begin{center}
Tianqing Liu$^{a}$, Xiaohui Yuan$^{b}$, Zhaohai Li$^{c}$ and  Aiyi Liu$^{d}$
\end{center}
\begin{center}

$^\texttt{a}${\it \small{School of Mathematics, Jilin University, Changchun, China}}\\
$^\texttt{b}${\it \small{School of Basic Science, Changchun University of Technology, Changchun, China}}
$^\texttt{c}${\it \small{Department of Statistics, George Washington University, USA}}\\
$^\texttt{d}${\it \small{Biostatistics and Bioinformatics Branch, Eunice Kennedy Shriver National Institute of Child Health and Human Development, Rockville, MD, USA}}

\end{center}

\footnotetext{$^*$Corresponding author.} \footnotetext{$\emph{\
Email\ addresses:} $\texttt{tianqingliu@outlook.com} (Tianqing Liu); \texttt{xiaohuiyuan@msn.com} (Xiaohui Yuan); \texttt{zli@gwu.edu} (Zhaohai Li); \texttt{liua@mail.nih.gov} (Aiyi Liu)}

\vskip0.2cm \noindent{\normalsize\bf Abstract:}  This paper considers an empirical likelihood inference for parameters defined by general estimating equations, when data are missing at random. The efficiency of existing estimators depends critically on correctly specifying the conditional expectation of the estimating function given the observed components of the random observations. When the conditional expectation is not correctly specified, the efficiency of estimation can be severely compromised even if the propensity function (of missingness) is correctly specified.  We propose an efficient estimator which enjoys the double-robustness property and can achieve the semiparametric efficiency bound within the class of the estimating functions that are generated by the estimating function of estimating equations, if both the propensity model and the regression model (of the conditional expectation) are specified correctly. Moreover, if the propensity model is specified correctly but the regression model is misspecified, the proposed estimator still achieves a semiparametric efficiency lower bound within a more general class of estimating functions. Simulation results suggest that the proposed estimators are robust against misspecification of the propensity model or regression model and outperform many existing competitors in the sense of having smaller mean-square errors. Moreover, using our approach for statistical inference requires neither resampling nor kernel smoothing. A real data example is used to illustrate the proposed approach.

\vskip0.2cm \noindent {\normalsize\bf Keywords:} Doubly robust inference; Empirical likelihood; Estimating equations; Generalized moment method; Missing data

\vskip 0.2cm
\section {Introduction} \label{sec1} \setcounter {equation}{0}
\def\theequation{\thesection.\arabic{equation}}
In medical and social science studies, estimating equations (EEs) (Boos, 1992; Godambe, 1991; Hansen, 1982; Qin \& Lawless, 1994) with missing data are often encountered and accommodate a wide range of data structure and parameters. Let $z=(x^\textsf{T},y^\textsf{T})^\textsf{T}$ be a vector of all
modelling variables. Suppose that we have a random sample of
incomplete data
\begin{eqnarray}
t_i=(x_i^\textsf{T},y_i^\textsf{T},\delta_i)^\textsf{T},\ \ i=1,2,\cdots,n, \label{data}
\end{eqnarray}
where all the $x_i$'s are observed, $\delta_i=0$ if $y_i$ is
missing, and otherwise $\delta_i=1$. Let $\beta^*$ be a $p$-dimensional parameter of interest in a model specified by the moment condition
\begin{eqnarray}
E\{s(z,\beta^*)\}=0, \label{gee}
\end{eqnarray}
where $s(z,\beta)=(s_1(z,\beta),\cdots,s_r(z,\beta))^\textsf{T}$ represents $r$ estimating functions for some $r\geq p$. We are interested in the inference about $\beta^*$ with the incomplete data (\ref{data}). Throughout, we make the missing-at-random (MAR) assumption, which implies that $\delta$ and $y$ are conditionally independent given $x$. As a result,
\begin{eqnarray}\label{mar}
P(\delta=1|y,x)=P(\delta=1|x):=\omega(x).
\end{eqnarray}
For a more detailed discussion on data missing completely at random (MCAR), MAR and non-ignorable missingness,
we refer readers to Rubin (1976) and Little \& Rubin (2002). A simple way to deal with missing data is to use only those data with complete observations. This method is known as complete-case analysis (CCA). The CCA method may result in a loss of efficiency, or, more seriously, biased results if missingness is not completely at random. To obtain valid inference, various debias methods have been studied in the literature, particularly the weighting method that was motivated by Horvitz and Thompson's (HT) (1952) estimators. To improve efficiency, Robins et al. (1994) proposed augmented EEs by using the parametrically estimated propensity scores to weight EEs. A well-known feature of the augmented method is that it can produce `doubly robust' inference, that is, the Robins-Rotnitzky-Zhao (RRZ) estimator is asymptotically unbiased if either the propensity model for $\omega(x)$ or regression model for $E\{s(x,y,\beta)|x\}$ is correctly specified. The RRZ estimator can achieve semiparametric efficiency if both the propensity model for $\omega(x)$ or regression model for $E\{s(x,y,\beta)|x\}$ are correctly specified, but is much less efficient if the regression model for $E\{s(x,y,\beta)|x\}$ is not close to $E\{s(x,y,\beta)|x\}$. The semiparametric efficiency bound for the parameter estimation in EEs models with missing data was studied in Chen, Hong \& Tarozzi (2008). By applying a sieve least-squares method,  they obtained semiparametric efficient estimators.

Empirical likelihood (EL) is a useful tool for finding estimators, constructing confidence regions, and testing hypotheses.
Some pioneering work on the EL method can be found in Owen (1988) and Qin (1994). The book by Owen (2001)
contains a comprehensive account of developments in EL. Kernel-assisted EL approaches have been employed to estimate $\beta^*$ in EEs (\ref{gee}) with incomplete data (\ref{data}). Zhou et al. (2008) proposed projecting the estimating equations containing missing data to the space generated by the observed data. Wang \& Chen (2009) proposed imputing missing data repeatedly from the estimated conditional distribution to remove the selection bias in the missingness. The estimators of Wang \& Chen (2009) and Zhou et al. (2008) have the same limiting covariance matrix, but are not semiparametric efficient in the sense of Chen, Hong \& Tarozzi (2008).  Tang \& Qin (2012) proposed augmenting nonparametric inverse probability weighted EEs and obtained an semiparametric efficient estimator by applying the EL approach to the augmented EEs. However, since kernel smoothing is used, the three approaches may be challenging for problems with high-dimensional non-missing variables due to the well known `curse of dimensionality'. To avoid the calculation when the dimensionality of covariate vector is high, Luo \& Wang (2015) proposed two estimators of the parameter vector defined by EEs in the presence of missing responses using weighted generalized method of moment (GMM) with the weights derived by EL under a dimension reduction constraint.  Chen, Leung \& Qin (2008) proposed an estimator of $\beta^*$ which is a solution to a set of weighted score equations by using EL weights obtained by leveraging the information that is contained in covariates and a surrogate outcome. Chen, Leung  and  Qin's estimator always gains in efficiency over the HT estimator. However, Luo \& Wang's approach and Chen, Leung \& Qin's approach assumes that the underlying missing data mechanism is either known or can be correctly specified.

EL approaches have also been applied to seek efficient and robust estimators of mean response with the assumption that data are missing at random. See, for instance, Wang \& Rao (2002), Wang et al. (2004) and Qin et al. (2009). In particular, Qin \& Zhang (2007) proposed a constrained EL estimation of mean response and showed that EL-based estimators enjoy the double-robustness property and can produce asymptotically unbiased and efficient estimators even if the true regression function is not completely known. Qin et al. (2008) further proposed a doubly robust regression imputation method and showed that asymptotically the sample mean based on the doubly robust regression imputation achieves the semiparametric efficiency bound if both regression and propensity models are specified correctly. However, it is not known whether the estimators of mean response in Qin \& Zhang (2007) and Qin et al. (2008) have smaller asymptotic variance than those of the HT and RRZ estimators.

Qin et al. (2009) proposed a unified EL approach for conditional mean model with partially missing covariates and explored the use of EL to effectively combine the complete data unbiased estimating equations and incomplete data unbiased estimating equations. They showed that the regression parameter estimator achieves the semiparametric efficiency bound in the sense of Bickel et al. (1993) if the propensity model is specified correctly. However, Qin, Zhang \& Leung's approach also assumes that the underlying missing data mechanism is either known or can be correctly specified.

In this paper, we propose an efficient and doubly robust (EDR) EL approach for making inference about $\beta^*$ in EEs (\ref{gee}) with incomplete data (\ref{data}). This approach efficiently incorporates the incomplete data into the data analysis by combining the complete data unbiased estimating equations and incomplete data unbiased estimating equations. The proposed EDR estimators enjoy the double-robustness property and can achieve the semiparametric efficiency bound in the sense of Chen, Hong \& Tarozzi (2008) if both the propensity model for $\omega(x)$ and regression model for $E\{s(x,y,\beta)|x\}$ are correctly specified. Simulation results suggest that the EDR estimators are robust to a misspecification of the propensity model for $\omega(x)$ or regression model for $E\{s(x,y,\beta)|x\}$ and outperform many existing competitors in the sense of having smaller mean-square errors. One important feature of our approach is that, it requires neither resampling nor kernel smoothing.

In addition, we extend the EL approach in Qin et al. (2009) to the case of EEs (\ref{gee}) with incomplete data (\ref{data}) and propose an EL estimator
which has the same asymptotic variance as the EDR estimator if the propensity model for $\omega(x)$ is specified correctly. However, it is difficult to compute the EL estimator due to the large number of estimating equations. Since the EL estimator is asymptotically efficient in the sense of Bickel et al. (1993) and has the same asymptotic covariance matrix as the EDR estimator, the EDR estimator is asymptotically efficient and thus more efficient than the HT and RRZ estimators if the propensity model for $\omega(x)$ is specified correctly.

The remainder of the paper is organized as follows. In Section 2, we extend the the weighting method (Horvitz \& Thompson, 1952) and augmented method (Robins et al., 1994) to estimation in EEs models (\ref{gee}) with incomplete data (\ref{data}). In Section 3, we propose our EDR approach and give the asymptotic properties of the EDR estimator.
In Section 4, we extend the EL approach in Qin et al. (2009) to the case of EEs (\ref{gee}) with incomplete data (\ref{data}) and propose the EL estimator and its asymptotic properties. Simulation results are given in Section 5. In Section 6, a real data example is used to illustrate the proposed EDR approach, and we conclude our paper in Section 7. The proofs of all forthcoming results
are postponed to the appendix.

\section{Unbiased estimating equations in missing data problems}
\setcounter {equation}{0}\def\theequation{\thesection.\arabic{equation}}
In this section, the non-missing-data probability $\omega(x)$ and $E\{s(z,\beta)|x\}$ are modeled parametrically and estimated from the observed data under the MAR assumption. Based on the estimators of $\omega(x)$ and $E\{s(z,\beta)|x\}$, we extend the the weighting method (Horvitz \& Thompson, 1952) and augmented method (Robins et al., 1994) to the case of EEs (\ref{gee}) with incomplete data (\ref{data}) and construct a class of unbiased estimating equations.

Let $\pi(x,\gamma)$ be a specified probability distribution function for given $\gamma$, a $q \times 1$ unknown vector parameter. According to White (1982), we define
\begin{eqnarray}\label{gammastar}
\gamma^*=\arg\max_{\gamma}\int \omega(x)\log\{\pi(x,\gamma)\}+\{1-\omega(x)\}\log\{1-\pi(x,\gamma)\}dF_X(x),
\end{eqnarray}
where $F_{X}(x)$ is the distribution function of $x$. It is natural to estimate $\gamma^*$ by the binomial likelihood estimator $\hat{\gamma}$ which is the solution of the following estimating equations
\begin{eqnarray}\label{u1}
\sum_{i=1}^nU_{1}(t_i,\gamma):=\sum_{i=1}^n\frac{\{\delta_i-\pi(x_i,\gamma)\}\partial\pi(x_i,\gamma)/\partial\gamma}{\pi(x_i,\gamma)\{1-\pi(x_i,\gamma)\}}=0.
\end{eqnarray}
When data is MAR, the commonly used HT estimator $\hat{\beta}_{HT}$ of $\beta^*$ is the solution to
\begin{eqnarray}
n^{-1} \sum_{i=1}^n\hat{V}_1(\beta,\hat{\gamma})\varphi_{1}(t_i, \beta,\hat{\gamma})=0,
\label{uht}
\end{eqnarray}
where
\begin{eqnarray}
\varphi_{1}(t_i, \beta,\gamma)=\delta_i s(z_i,\beta)/\pi(x_i,\gamma)
\label{psi1}
\end{eqnarray}
with
\begin{eqnarray*}
&&\hat{V}_1(\beta,\gamma)=\hat{V}_{12}^\textsf{T}(\beta,\gamma)\hat{V}_{11}^{-1}(\beta,\gamma)\\
&&\hat{V}_{11}(\beta,\gamma)=n^{-1}\sum_{i=1}^n \{\varphi_{1}(t_i, \beta,\gamma)-\hat{A}_{1}(\beta,\gamma)\hat{A}_{2}^{-1}(\gamma)U_{1}(t_i,\gamma)\}^{\otimes2},\\
&&\hat{V}_{12}(\beta,\gamma)=n^{-1}\sum_{i=1}^n \partial\varphi_{1}(t_i, \beta,\gamma)/\partial\beta^\textsf{T}, \\
&&\hat{A}_{1}(\beta,\gamma)=n^{-1}\sum_{i=1}^n\partial \varphi_{1}(t_i, \beta,\gamma)/\partial\gamma^\textsf{T},\ \ \hat{A}_{2}(\gamma)=n^{-1}\sum_{i=1}^n\partial U_{1}(t_i,\gamma) / \partial  \gamma^\textsf{T},
\end{eqnarray*}
and for a vector $e$, $e^{\otimes2}=ee^\textsf{T}$. By augmenting the HT estimating equations in (\ref{uht}), Robins, Rotnitzky \& Zhao (1994) proposed estimating $\beta^*$ using $\hat{\beta}_{RRZ}$, which is obtained by solving
\begin{eqnarray}
\label{urrz}
n^{-1}\sum_{i=1}^n \hat{V}_2(\beta,\hat{\gamma},\hat{\alpha})\varphi_{2}(t_i,\beta,\hat{\gamma},\hat{\alpha})=0,
\end{eqnarray}
where
\begin{eqnarray}
\varphi_{2}(t_i,\beta,\gamma,\alpha)= \frac{\delta_i}{\pi(x_i,\gamma)}s(z_i,\beta)-\frac{\delta_i-\pi(x_i,\gamma)}{\pi(x_i,\gamma)}u(x_i,\beta,\alpha)
\label{psi2}
\end{eqnarray}
and
\begin{eqnarray*}
&&\hat{V}_2(\beta,\gamma,\alpha)=\hat{V}_{22}^\textsf{T}(\beta,\gamma,\alpha)\hat{V}_{21}^{-1}(\beta,\gamma,\alpha)\\
&&\hat{V}_{21}(\beta,\gamma,\alpha)=n^{-1}\sum_{i=1}^n \{\varphi_{2}(t_i, \beta,\gamma,\alpha)-\hat{B}_{1}(\beta,\gamma,\alpha)\hat{B}_{2}^{-1}(\gamma,\alpha)U_{12}(t_i,\gamma,\alpha)\}^{\otimes2},\\
&&\hat{V}_{22}(\beta,\gamma,\alpha)=n^{-1}\sum_{i=1}^n \partial\varphi_{2}(t_i, \beta,\gamma,\alpha)/\partial\beta^\textsf{T},\ \ U_{12}(t_i,\gamma,\alpha)=(U_{1}^\textsf{T}(t_i,\gamma),U_{2}^\textsf{T}(t_i,\alpha))^\textsf{T}, \\
&&\hat{B}_{1}(\beta,\gamma,\alpha)=n^{-1}\sum_{i=1}^n\frac{\partial \varphi_{2}(t_i, \beta,\gamma,\alpha) }{\partial(\gamma^\textsf{T},\alpha^\textsf{T})},\ \ \hat{B}_2(\gamma,\alpha)=n^{-1}\sum_{i=1}^n\frac{\partial U_{12}(t_i,\gamma,\alpha) }{\partial  (\gamma^\textsf{T},\alpha^\textsf{T})}.
\end{eqnarray*}
Here, $u(\cdot)$ is a $r\times1$ vector of known functions of $x$, up to the unknown parameter $\beta$ and another unknown (vector) parameter $\alpha$. The optimal choice for $u(\cdot)$ is given by $E\{s(z,\beta)|x\}$. Since $E\{s(z,\beta)|x\}$ is unknown, it needs to be estimated using the observed data. One popular approach is to fit a flexible conditional distribution model for $f_{Y|X}(y|x)$, which is the conditional density or probability function of $y$ given $x$. Let $f(y|x,\alpha)$ be a working model for $f_{Y|X}(y|x)$. Then, a working model for $E\{s(z,\beta)|x\}$ is given by $u(x,\beta,\alpha)=\int s(z,\beta)f(y|x,\alpha)dy$. According to White (1982), we define
\begin{eqnarray}\label{aphstar}
\alpha^*=\arg\max_{\alpha}\int\int \omega(x)f_{Y|X}(y|x)\log\{f(y|x,\alpha)\}dydF_X(x).
\end{eqnarray}
Then, $\alpha^*$ can be estimated by the conditional likelihood estimator $\hat{\alpha}$ which maximizes the conditional log-likelihood $\sum_{i=1}^n\delta_i\log\{f(y_i|x_i,\alpha)\}$. Obviously, $\hat{\alpha}$ satisfies the following estimating equations
\begin{eqnarray}\label{u2}
\sum_{i=1}^nU_{2}(t_i,\alpha):=\sum_{i=1}^n\delta_i\frac{\partial\log\{f(y_i|x_i,\alpha)\}}{\partial\alpha}=0.
\end{eqnarray}
The asymptotic distribution of $\hat{\beta}_{HT}$ and $\hat{\beta}_{RRZ}$ can be derived similarly to that of $\hat{\beta}_{EDR}$ in theorem \ref{elw0}. The following theorems summarize the large-sample results of $\hat{\beta}_{HT}$ and $\hat{\beta}_{RRZ}$.
\begin{theorem}\label{HT}
Suppose $\pi(x,\gamma^*)=\omega(x)$. Under suitable conditions,  $n^{1/2}(\hat{\beta}_{HT}-\beta^*)\drow N(0,\Sigma_{HT})$ as $n\rightarrow\infty$, where \begin{eqnarray*}
\Sigma_{HT}
&=&\left\{E\left(\frac{\partial\varphi_1^\textsf{T}}{\partial\beta}\right)\left(E\left[\varphi_1-E\left(\frac{\partial\varphi_1}{\partial\gamma^\textsf{T}}\right)\left\{E\left(\frac{\partial U_{1}}{\partial\gamma^\textsf{T}}\right)\right\}^{-1}U_{1}\right]^{\otimes2}\right)^{-1}E\left(\frac{\partial\varphi_1}{\partial\beta^\textsf{T}}\right)\right\}^{-1}\\
&=&\{F_\beta^\textsf{T}(S_{\varphi_1}-F_\gamma S_B^{-1}F_\gamma^\textsf{T})^{-1}F_\beta\}^{-1},
\end{eqnarray*}
$\varphi_{1}=\varphi_{1}(t, \beta^*,\gamma^*)$, $U_{1}=U_{1}(t,\gamma^*)$, $E\left(\frac{\partial U_1}{\partial\gamma^\textsf{T}}\right)=E\left\{\frac{\partial U_{1}(t,\gamma^*)}{\partial\gamma^\textsf{T}}\right\}$, $F_\beta=E\left(\frac{\partial\varphi_1}{\partial\beta^\textsf{T}}\right)=E\left\{\frac{\partial\varphi_{1}(t, \beta^*,\gamma^*)}{\partial\beta^\textsf{T}}\right\}$, $F_\gamma=E\left(\frac{\partial\varphi_1}{\partial\gamma^\textsf{T}}\right)=E\left\{\frac{\partial\varphi_{1}(t, \beta^*,\gamma^*)}{\partial\gamma^\textsf{T}}\right\}$,
$S_{\varphi_1}=E(\varphi_1^{\otimes2})$ and $S_B=E(U_1^{\otimes2})$.
\end{theorem}
\begin{theorem}\label{RRZ}
Suppose $\pi(x,\gamma^*)=\omega(x)$ or $u(x,\beta,\alpha^*)=E\{s(z,\beta)|x\}$. Under suitable conditions,  $n^{1/2}(\hat{\beta}_{RRZ}-\beta^*)\drow N(0,\Sigma_{RRZ})$ as $n\rightarrow\infty$, where
\begin{eqnarray*}
\Sigma_{RRZ}&=&\left\{E\left(\frac{\partial\varphi_2^\textsf{T}}{\partial\beta}\right)\left(E\left[\varphi_2-E\left(\frac{\partial\varphi_2}{\partial\gamma^\textsf{T}}\right)\left\{E\left(\frac{\partial U_{1}}{\partial\gamma^\textsf{T}}\right)\right\}^{-1}U_{1}\right.\right.\right.\\
&&-\left.\left.\left.E\left(\frac{\partial\varphi_2}{\partial\alpha^\textsf{T}}\right)\left\{E\left(\frac{\partial U_{2}}{\partial\alpha^\textsf{T}}\right)\right\}^{-1}U_{2}\right]^{\otimes2}\right)^{-1}E\left(\frac{\partial\varphi_2}{\partial\beta^\textsf{T}}\right)\right\}^{-1},
\end{eqnarray*}
$\varphi_2=\varphi_{2}(t,\beta^*,\gamma^*,\alpha^*)$, $U_1=U_{1}(t,\gamma^*)$, $U_2=U_{2}(t,\alpha^*)$, $E\left(\frac{\partial U_1}{\partial\gamma^\textsf{T}}\right)=E\left\{\frac{\partial U_{1}(t,\gamma^*)}{\partial\gamma^\textsf{T}}\right\}$, $E\left(\frac{\partial U_2}{\partial\alpha^\textsf{T}}\right)=E\left\{\frac{\partial U_{2}(t,\alpha^*)}{\partial\alpha^\textsf{T}}\right\}$, $E\left(\frac{\partial\varphi_2}{\partial\beta^\textsf{T}}\right)=E\left\{\frac{\partial\varphi_{2}(t,\beta^*,\gamma^*,\alpha^*)}{\partial\beta^\textsf{T}}\right\}$, $E\left(\frac{\partial\varphi_2}{\partial\gamma^\textsf{T}}\right)=E\left\{\frac{\partial\varphi_{2}(t,\beta^*,\gamma^*,\alpha^*)}{\partial\gamma^\textsf{T}}\right\}$ and
$E\left(\frac{\partial\varphi_2}{\partial\alpha^\textsf{T}}\right)=E\left\{\frac{\partial\varphi_{2}(t,\beta^*,\gamma^*,\alpha^*)}{\partial\alpha^\textsf{T}}\right\}$.
\end{theorem}
\begin{remark}\label{rrzbound}
If $\pi(x,\gamma^*)=\omega(x)$, we have $E\left(\frac{\partial\varphi_2}{\partial\alpha^\textsf{T}}\right)=0$. If $u(x,\beta,\alpha^*)=E\{s(z,\beta)|x\}$, we have $E\left(\frac{\partial\varphi_2}{\partial\gamma^\textsf{T}}\right)=0$. Suppose $\pi(x,\gamma^*)=\omega(x)$ and $u(x,\beta,\alpha^*)=E\{s(z,\beta)|x\}$, it follows that $\Sigma_{RRZ}=\Sigma_{0}$, where
\begin{eqnarray*}
\Sigma_{0}&=&\left[E\left(\frac{\partial{\varphi_2^\textsf{T}}}{\partial\beta}\right)\{E(\varphi_2^{\otimes2})\}^{-1}E\left(\frac{\partial\varphi_2}{\partial\beta^\textsf{T}}\right)\right]^{-1}\\
&=&\left(F_\beta^\textsf{T}\left[E\{\pi^{-1}(s-u)^{\otimes2}\}+E(u^{\otimes2})\right]^{-1}F_\beta\right)^{-1},
\end{eqnarray*}
is the semiparametric efficiency bound in the sense of Chen, Hong $\&$ Tarozzi (2008).
\end{remark}
Theorem \ref{RRZ} allows us to construct a doubly robust estimator of the covariance matrix $\Sigma_{RRZ}$.
\begin{remark}\label{rrzvar}
Let $\hat{\Sigma}_{RRZ}=\{\hat{V}_{22}^\textsf{T}(\hat{\beta}_{RRZ},\hat{\gamma},\hat{\alpha})\hat{V}_{21}^{-1}(\hat{\beta}_{RRZ},\hat{\gamma},\hat{\alpha})\hat{V}_{22}(\hat{\beta}_{RRZ},\hat{\gamma},\hat{\alpha})\}^{-1}$. Then, $\hat{\Sigma}_{RRZ}$ is a consistent doubly robust estimator of $\Sigma_{RRZ}$. $\hat{\Sigma}_{RRZ}$ also can be used to construct a doubly robust confidence region for $\beta^*$.
\end{remark}
Notice that when the number of estimating equations and parameters are equal, the estimating equations in (\ref{uht}) and (\ref{urrz}) can be simplified as follows.
\begin{remark}\label{retp1}
When $r=p$, namely the number of estimating equations is the same as the dimension of $\beta^*$, we set $\hat{V}_1(\cdot)=\hat{V}_2(\cdot)=I_p$ in (\ref{uht}) and (\ref{urrz}), respectively. The conclusions in Theorems \ref{HT}-\ref{RRZ} and Remarks \ref{rrzbound}-\ref{rrzvar} hold.
\end{remark}

\section{Efficient and doubly robust EL estimator} \setcounter {equation}{0}
\def\theequation{\thesection.\arabic{equation}}
In this section, we employ the EL method to seek a constrained EL estimatior of $\beta^*$ with incomplete data (\ref{data}). For $i=1,\cdots,n$, write
\begin{eqnarray}
h(t,\beta,\gamma,\alpha)&=&\frac{\delta-\pi(x,\gamma)}{\pi(x,\gamma)}\xi(x,\beta,\alpha),\nonumber\\
\xi(x,\beta,\alpha)&=&(u^\textsf{T}(x,\beta,\alpha),a^\textsf{T}(x,\beta,\alpha))^\textsf{T},\label{xi}\\
g(t,\beta,\gamma,\alpha)&=&(h^\textsf{T}(t,\beta,\gamma,\alpha),U_{1}^\textsf{T}(t,\gamma))^\textsf{T},\nonumber
\end{eqnarray}
where $a(x,\beta,\alpha)$ is a vector of known functions of $x$, up to the unknown parameter $\beta$ and $\alpha$. To this end, let $p_i$ represent the probability weight allocated to $g(t_i,\beta,\gamma,\alpha)$. Then, we maximize the log-EL function $\sum_{i=1}^n\log p_i$ subject to the constraints
\begin{eqnarray*}
p_i\geq0,\ \ \ \sum_{i=1}^np_{i}=1,\ \ \
\sum_{i=1}^np_{i}g(t_i,\beta,\gamma,\alpha)=0.
\end{eqnarray*}
By using the Lagrange multiplier method,  we find that the optimal
$p_i$ is
\begin{eqnarray}
p(t_i,\beta,\theta)=\frac{1}{n}\frac{1}{1+\lambda^\textsf{T}g(t_i,\beta,\gamma,\alpha)},
\label{elpi}
\end{eqnarray}
where $\theta=(\gamma^\textsf{T},\alpha^\textsf{T},\lambda^\textsf{T})^\textsf{T}$ and $\lambda=\hat{\lambda}(\beta,\gamma,\alpha)$ is the Lagrange multiplier that satisfies
\begin{eqnarray}
\sum_{i=1}^nU_{3}(t_i,\beta,\theta):=\sum_{i=1}^n\frac{g(t_i,\beta,\gamma,\alpha)}{1+\lambda^\textsf{T}g(t_i,\beta,\gamma,\alpha)}=0.
\label{u3}
\end{eqnarray}
Let
\begin{eqnarray}\label{aphstar}
\lambda(\beta,\gamma,\alpha)=\arg\max_{\lambda}\int\int \log\{1+\lambda^\textsf{T}g(t_i,\beta,\gamma,\alpha)\}dF_{X,\delta}(x_i,\delta_i),
\end{eqnarray}
and $\lambda^*=\lambda(\beta^*,\gamma^*,\alpha^*)$, where $F_{X,\delta}(\cdot,\cdot)$ is the joint distribution function of $(x,\delta)$. Obviously, for fixed $(\beta,\gamma,\alpha)=(\beta^*,\gamma^*,\alpha^*)$, (\ref{u3}) is an unbiased estimating equations for $\lambda^*$. We define the EDR estimator, $\hat{\beta}_{EDR}$, of $\beta^*$ as the solution to
\begin{eqnarray}
\label{uel}
n^{-1} \sum_{i=1}^n\hat{V}_3(\beta,\hat{\theta}(\beta))\varphi_{3}(t_i,\beta,\hat{\theta}(\beta))=0,
\end{eqnarray}
where $\hat{\theta}(\beta)=(\hat{\gamma}^\textsf{T},\hat{\alpha}^\textsf{T},\hat{\lambda}^\textsf{T}(\beta,\hat{\gamma},\hat{\alpha}))^\textsf{T}$, \begin{eqnarray}\label{psi3}
\varphi_{3}(t_i,\beta,\theta)=\frac{\delta_i s(z_i,\beta)}{\pi(x_i,\gamma)\{1+\lambda^\textsf{T}g(t_i,\beta,\gamma,\alpha)\}}+\frac{u(x_i,\beta,\alpha)g^\textsf{T}(t_i,\beta,\gamma,\alpha)\lambda}{1+\lambda^\textsf{T}g(t_i,\beta,\gamma,\alpha)},
\end{eqnarray}
\begin{eqnarray*}
&&\hat{V}_3(\beta,\theta)=\hat{V}_{32}^\textsf{T}(\beta,\theta)\hat{V}_{31}^{-1}(\beta,\theta)\\
&&\hat{V}_{31}(\beta,\theta)=n^{-1}\sum_{i=1}^n \{\varphi_{3}(t_i, \beta,\theta)-\hat{C}_{1}(\beta,\theta)\hat{C}_{2}^{-1}(\beta,\theta)U(t_i,\beta,\theta)\}^{\otimes2},\\
&&\hat{V}_{32}(\beta,\theta)=n^{-1}\sum_{i=1}^n\left\{\partial\varphi_3{(t_i, \beta,\theta)/\partial\beta^\textsf{T}}-\hat{C}_{1}(\beta,\theta)\hat{C}_{2}^{-1}(\beta,\theta)\partial{U}(t_i, \beta,\theta)/\partial\beta^\textsf{T}\right\},\\ &&\hat{C}_{1}(\beta,\theta)=n^{-1}\sum_{i=1}^n\partial \varphi_{3}(t_i, \beta,\theta) / \partial  \theta^\textsf{T},\ \ \hat{C}_2(\beta,\theta)=n^{-1}\sum_{i=1}^n\partial U(t_i,\beta,\theta) / \partial  \theta^\textsf{T},
\end{eqnarray*}
and
\begin{eqnarray}\label{u1234}
U(t_i,\beta,\theta)&=&(U_{1}^\textsf{T}(t_i,\gamma),U_{2}^\textsf{T}(t_i,\alpha),U_{3}^\textsf{T}(t_i,\beta,\theta))^\textsf{T}.
\end{eqnarray}
It is easily seen that
\begin{eqnarray*}
&&n^{-1}\sum_{i=1}^n\varphi_{3}(t_i,\beta,\theta)\\
&=&\sum_{i=1}^np(t_i,\beta,\theta)\left\{\frac{\delta_i s(z_i,\beta)}{\pi(x_i,\gamma)}\right\}
-n^{-1}\sum_{i=1}^n\{np(t_i,\beta,\theta)-1\}u(x_i,\beta,\alpha)\\
&=&\sum_{i=1}^np(t_i,\beta,\theta)\left\{\frac{\delta_i s(z_i,\beta)}{\pi(x_i,\gamma)}\right\}
+n^{-1}\sum_{i=1}^nu(x_i,\beta,\alpha)-\sum_{i=1}^np(t_i,\beta,\theta)u(x_i,\beta,\alpha)\\
&=&n^{-1}\sum_{i=1}^nu(x_i,\beta,\alpha)+\sum_{i=1}^np(t_i,\beta,\theta)\left[\frac{\delta_i \{s(z_i,\beta)-u(x_i,\beta,\alpha)\}}{\pi(x_i,\gamma)}\right].
\end{eqnarray*}
This expression gives some intuitive insight to the doubly robustness of $\hat{\beta}_{EDR}$. Let $\vartheta=(\beta^\textsf{T},\theta^\textsf{T})^\textsf{T}$, $\vartheta^*=(\beta^{*\textsf{T}},\theta^{*\textsf{T}})^\textsf{T}$ and $\theta^*=(\gamma^{*\textsf{T}},\alpha^{*\textsf{T}},\lambda^{*\textsf{T}})^\textsf{T}$. The following theorem summarizes the large-sample results of $\hat{\beta}_{EDR}$.
\begin{theorem}\label{elw0}
Assume $\pi(x,\gamma^*)=\omega(x)$ or $u(x,\beta,\alpha^*)=E\{s(z,\beta)|x\}$, that is, the propensity model $\pi(x,\gamma)$ or the regression model $u(x,\beta,\alpha)$ is correctly specified. Under regularity conditions in the Appendix,  $n^{1/2}(\hat{\beta}_{EDR}-\beta^*)\drow N(0,\Sigma_{EDR})$ as $n\rightarrow\infty$, where $\Sigma_{EDR}=\{V_{32}^\textsf{T}(\vartheta^*)V_{31}^{-1}(\vartheta^*)V_{32}(\vartheta^*)\}^{-1}$, where
\begin{eqnarray*}\label{Sigma0}
V_{31}(\vartheta)&=&E\left(\left[\varphi_{3}(t, \vartheta)-E\left\{\frac{\partial \varphi_{3}(t, \vartheta)}{\partial  \theta^\textsf{T}} \right\} \left(E\left\{\frac{\partial U(t,\vartheta)}{\partial  \theta^\textsf{T}}\right\}\right)^{-1}U(t,\vartheta)\right]^{\otimes2}\right),\\
V_{32}(\vartheta)&=&E\left\{\frac{\partial\varphi_3(t, \vartheta)}{\partial\beta^\textsf{T}}\right\}-E\left\{\frac{\partial \varphi_{3}(t, \vartheta)}{\partial  \theta^\textsf{T}} \right\} \left(E\left\{\frac{\partial U(t,\vartheta)}{\partial  \theta^\textsf{T}}\right\}\right)^{-1}E\left\{\frac{\partial{U}(t, \vartheta)}{\partial\beta^\textsf{T}}\right\},
\end{eqnarray*}
and $\varphi_{3}(\cdot)$ and $U(\cdot)$ are defined in (\ref{psi3}) and (\ref{u1234}), respectively.
\end{theorem}
Using theorem \ref{elw0}, the asymptotic distribution of $\hat{\beta}_{EDR}$ can be obtained in the case of correctly specified propensity function $\pi(x,\gamma)$ but arbitrary conditional expectation function $u(x,\beta,\alpha)$.
We also write
\begin{eqnarray}
&&F_h=E\{\varphi_1(t,\beta^*,\gamma^*)h^\textsf{T}(t,\beta^*,\gamma^*,\alpha^*)\},
\nonumber \\
&&F_g=E\{\varphi_1(t,\beta^*,\gamma^*)g^\textsf{T}(t,\beta^*,\gamma^*,\alpha^*)\},
\nonumber \\
&&S_h=E\{h(t,\beta^*,\gamma^*,\alpha^*)h^\textsf{T}(t,\beta^*,\gamma^*,\alpha^*)\},
\label{nota}\\
&&S_g=E\{g(t,\beta^*,\gamma^*,\alpha^*)g^\textsf{T}(t,\beta^*,\gamma^*,\alpha^*)\},
\nonumber \\
&&H_\gamma=E\left\{\frac{\partial
h(t,\beta^*,\gamma^*,\alpha^*)}{\partial \gamma^\textsf{T}}\right\}.
\nonumber
\end{eqnarray}
\begin{theorem}\label{influence}
Assume $\pi(x,\gamma^*)=\omega(x)$, that is, the propensity model $\pi(x,\gamma)$ is correctly specified, we have $\Sigma_{EDR}=\Sigma_1$, where
\begin{eqnarray*}
&&\Sigma_1^{-1}\\
&=&F_\beta^\textsf{T}\left\{E\left(\left[\varphi_1(t,\beta^*,\gamma^*)-F_gS_g^{-1}g(t,\beta^*,\gamma^*,\alpha^*)\right]^{\otimes2}\right)\right\}^{-1}F_\beta\\
&=&F_{\beta}^\textsf{T}(S_{\varphi_1}-F_gS_g^{-1}F_g^\textsf{T})^{-1}F_{\beta}\\
&=&F_\beta^\textsf{T}\{S_{\varphi_1}-F_\gamma S_B^{-1}F_\gamma^\textsf{T}-(F_\gamma
S_B^{-1}H_{\gamma}^\textsf{T}-F_h)(S_h-H_{\gamma}S_B^{-1}H_{\gamma}^\textsf{T})^{-1}(F_\gamma
S_B^{-1}H_{\gamma}^\textsf{T}-F_h)^\textsf{T}\}^{-1}F_\beta.
\end{eqnarray*}
\end{theorem}
\begin{remark}\label{afun}
Theorem \ref{el} shows that $\hat{\beta}_{EDR}$ is asymptotically efficient within the class of estimating functions that are generated by $\varphi_{1}(t, \beta,\gamma)$ and $g(t,\beta,\gamma,\alpha)$ when the propensity model $\pi(x,\gamma)$ is correctly specified. In theory, including more estimating functions, say $a(x,\beta,\alpha)$ in (\ref{xi}), leads to more efficient estimator, asymptotically (Corollaries 1-2 in Qin $\&$ Lawless (1994)). However, in finite samples, including too many estimating functions that are not sensitive to the unknown parameter may actually hurt efficiency. Thus, unless mentioned otherwise, we set $a(x,\beta,\alpha)\equiv1$.
\end{remark}
The following theorem establishes the asymptotic equivalence of $\hat{\beta}_{EDR}$ and $\hat{\beta}_{RRZ}$ when $\pi(x,\gamma)$ and $u(x,\beta,\alpha)$ are both correctly specified.
\begin{theorem}\label{elwbound}
Assume $\pi(x,\gamma^*)=\omega(x)$ and $u(x,\beta,\alpha^*)=E\{s(z,\beta)|x\}$. Then, $\Sigma_{EDR}=\Sigma_0$,
which means that $\hat{\beta}_{EDR}$ achieves semiparametric full efficiency in the sense of Chen, Hong $\&$ Tarozzi (2008).
\end{theorem}
Theorem \ref{elw0} allows us to construct a doubly robust estimator of the covariance matrix $\Sigma_{EDR}$.
\begin{remark}\label{elvar}
Let $\hat{\Sigma}_{EDR}=\{\hat{V}_{32}^\textsf{T}(\hat{\vartheta})\hat{V}_{31}^{-1}(\hat{\vartheta})\hat{V}_{32}(\hat{\vartheta})\}^{-1}$ with $\hat{\vartheta}=(\hat{\beta}_{EDR}^\textsf{T},\hat{\theta}^\textsf{T})^\textsf{T}$ and $\hat{\theta}=\hat{\theta}(\hat{\beta}_{EDR})=(\hat{\gamma}^\textsf{T},\hat{\alpha}^\textsf{T},\hat{\lambda}^\textsf{T}(\hat{\beta}_{EDR},\hat{\gamma},\hat{\alpha}))^\textsf{T}$.
Then, $\hat{\Sigma}_{EDR}$ is a consistent doubly robust estimator of $\Sigma_{EDR}$. $\hat{\Sigma}_{EDR}$ also can be used to construct a doubly robust confidence region for $\beta^*$.
\end{remark}
Notice that when the number of estimating equations and parameters are equal, the estimating equations in (\ref{uel}) can be simplified as follows.
\begin{remark}\label{retp2}
When $r=p$, namely the number of estimating equations is the same as the dimension of $\beta^*$, we set $\hat{V}_3(\cdot)=I_p$ in (\ref{uel}). The conclusions in Theorems \ref{elw0}-\ref{elwbound} and Remarks \ref{afun}-\ref{elvar} hold.
\end{remark}

\begin{remark}\label{response}
A natural application of the proposed procedure is the estimation of the mean response. Denote the response variable and covariate vector as $y$ and $x$. Let $s(z,\beta^*)=y-\beta^*$ in equation (\ref{gee}). Then, $\beta^*$ is the mean response and
\begin{eqnarray}
\hat{\beta}_{EDR}&=&\sum_{i=1}^n\hat{p}_i\left\{\frac{\delta_i y_i}{\pi(x_i,\hat{\gamma})}\right\}
-n^{-1}\sum_{i=1}^n\{n\hat{p}_i-1\}m(x_i,\hat{\alpha}),
\label{edrmean}
\end{eqnarray}
where $m(x,\hat{\alpha})$ is the regression model for $E(y|x)$, $\hat{p}_i=n^{-1}\{1+\hat{\lambda}^\textsf{T}g(t_i,\hat{\gamma},\hat{\alpha})\}^{-1}$ and $\hat{\lambda}$ is the Lagrange multiplier that satisfies $\sum_{i=1}^ng(t_i,\hat{\gamma},\hat{\alpha})/\{1+\lambda^\textsf{T}g(t_i,\hat{\gamma},\hat{\alpha})\}=0$. Here, $g(t,\gamma,\alpha)=(h^\textsf{T}(t,\gamma,\alpha),U_{1}^\textsf{T}(t,\gamma))^\textsf{T}$, $h(t,\gamma,\alpha)=\{\delta-\pi(x,\gamma)\}\xi(x,\alpha)/\pi(x,\gamma)$ and $\xi(x,\alpha)=(m(x,\alpha),1)^\textsf{T}$.
\end{remark}

\section{EL estimation of $(\beta,\gamma)$ when $\pi(x,\gamma^*)=\omega(x)$}

In this section, we extend the EL approach (Qin \& Lawless, 1994; Qin et al., 2009) to EEs (\ref{gee}) with incomplete data (\ref{data}) and propose the EL estimator of $(\beta,\gamma)$. Write
\begin{eqnarray*}\label{psifun}
\psi(t,\beta,\gamma)&=&(\varphi_1^\textsf{T}(t,\beta,\gamma),g^\textsf{T}(t,\beta,\gamma,\hat{\alpha}))^\textsf{T}.
\end{eqnarray*}
For the sake of parsimony, we suppress $\hat{\alpha}$ from the estimating function $\psi(t,\beta,\gamma)$ since the large sample results for the EL estimation of $(\beta,\gamma)$ are unaffected by $\hat{\alpha}$ when $\pi(x,\gamma^*)=\omega(x)$. Notice that the dimension of $\psi(t,\beta,\gamma)$ is higher than that of $(\beta,\gamma)$, one may employ the profile EL method (Qin \& Lawless, 1994; Qin et al., 2009) to seek an optimal combination of the estimating functions $\psi(t_i,\beta,\gamma)$.
To this end, let $L_{EL}=\prod_{i=1}^np_i$, where $p_i$, $i=1,\cdots,n$, are nonnegative jump sizes with total mass that
sums to 1. For fixed $(\beta,\gamma)$, we maximize $L_{EL}$ subject to the constraints
\begin{eqnarray*}
p_i\geq0,\ \ \ \sum_{i=1}^np_{i}=1,\ \ \
\sum_{i=1}^np_{i}\psi(t_i,\beta,\gamma)=0.
\end{eqnarray*}
After profiling the $p_i$'s, the profile empirical log-likelihood of $(\beta,\gamma)$ is given by
\begin{eqnarray}
\ell_{EL}(\beta,\gamma)=-\sum_{i=1}^n\log\{1+\mu^\textsf{T}\psi(t_i,\beta,\gamma)\}-n\log n,
\label{ell}
\end{eqnarray}
where $\mu=\mu(\beta,\gamma)$ is determined by
\begin{eqnarray*}
\frac{1}{n}\sum_{i=1}^n\frac{\psi(t_i,\beta,\gamma)}{1+\mu^\textsf{T}\psi(t_i,\beta,\gamma)}=0.
\end{eqnarray*}
Let $(\hat{\beta}_{EL},\hat{\gamma}_{EL})$ denote the EL estimator of $(\beta,\gamma)$ that maximizes $\ell_{EL}(\beta,\gamma)$. The following theorem summarizes the large-sample results of $(\hat{\beta}_{EL},\hat{\gamma}_{EL})$.

\begin{theorem}\label{el}
Suppose the missing-data mechanism $\pi(x,\gamma)$ is correctly specified and $\psi(t,\beta,\gamma)$ satisfies the regularity conditions in Theorem 1 of Qin and Lawless (1994), we have
\begin{eqnarray}
n^{1/2}\left(\begin{array}{c}
\hat{\beta}_{EL}-\beta^*\\
\hat{\gamma}_{EL}-\gamma^*\\
\end{array}\right)
\drow N\left(0,\left(\begin{array}{cc}
\Sigma_{1}&0\\
0&S_B^{-1}\\
\end{array}\right) \right) \label{elnormal}
\end{eqnarray}
as $n\rightarrow\infty$, where $\Sigma_{1}$ is defined in Theorem \ref{influence}.
\end{theorem}
Theorem \ref{el} demonstrate that $\hat{\beta}_{EL}$ and $\hat{\beta}_{EDR}$ are asymptotically equivalent when the propensity model $\pi(x,\gamma)$ is correctly specified. Also based on Corollary 2 in Qin and Lawless (1994), $\hat{\beta}_{EL}$ is the optimal estimator in the class of estimating functions that are linear combinations of $\psi(t_i,\beta,\gamma)$.
Note that $\hat{\beta}_{HT}$ can be written as the
solution to
$$
\sum_{i=1}^nW_{HT}(\beta,\gamma)\psi(t_i,\beta,\gamma)=0,\ \
W_{HT}(\beta,\gamma)=\left(\begin{array}{ccc}
\hat{V}_1(\beta,\gamma)&0_{p\times(p+1)}&0_{p\times q}\\
0_{q\times p}&0_{q\times(p+1)}&I_{q\times q}\\
\end{array}\right),
$$
which implies that the proposed estimators $\hat{\beta}_{EL}$ and $\hat{\beta}_{EDR}$ are
asymptotically more efficient than $\hat{\beta}_{HT}$. For two matrices $A$ and $B$, we write $A \leq B$ if $B-A$ is a nonnegative-definite matrix. In fact, Theorems \ref{HT} and \ref{influence} lead to the following result.
\begin{corollary}
If both $S_{\varphi_1}-F_\gamma S_B^{-1}F_\gamma^\textsf{T}$ and $S_h-H_{\gamma}S_B^{-1}H_{\gamma}^\textsf{T}$ are positive definite, we
have $\Sigma_{EDR}\leq \Sigma_{HT}$, and the equality holds if and only if
$F_\gamma S_B^{-1}H_{\gamma}^\textsf{T}=F_h$.
\end{corollary}
Similarly, $\hat{\beta}_{RRZ}$ is the solution to the equations
$$
\sum_{i=1}^nW_{RRZ}(\beta,\gamma)\psi(t_i,\beta,\gamma)=0,\ \
W_{RRZ}(\beta,\gamma)=\left(\begin{array}{cccc}
\hat{V}_2(\beta,\gamma,\hat{\alpha})&-\hat{V}_2(\beta,\gamma,\hat{\alpha})&0_{p\times1}&0_{p\times q}\\
0_{q\times p}&0_{q\times p}&0_{q\times1}&I_{q\times q}\\
\end{array}\right).
$$
As a result, the optimal estimators $\hat{\beta}_{EL}$ and $\hat{\beta}_{EDR}$ are asymptotically more efficient than $\hat{\beta}_{RRZ}$.

In conclusion, letting $\gg$ stand for ``asymptotically more efficient
than,'' we have the following relationships:
$\hat{\beta}_{EL}=\hat{\beta}_{EDR}\gg\hat{\beta}_{HT}$ and $\hat{\beta}_{EL}=\hat{\beta}_{EDR}\gg\hat{\beta}_{RRZ}$.
If $u(x,\beta,\alpha^*)=E\{s(z,\beta)|x\}$, we obtain $\hat{\beta}_{EL}=\hat{\beta}_{EDR}=\hat{\beta}_{RRZ}\gg\hat{\beta}_{HT}$.
In practice, a prudent choice of $u(x,\beta,\alpha)$ should lead to an estimator of $\beta^*$ that is more efficient than $\hat{\beta}_{HT}$ and $\hat{\beta}_{RRZ}$ when the missing-data mechanism $\pi(x,\gamma)$ is correctly specified, whereas no such guarantee can be said when the propensity model $\pi(x,\gamma)$ is misspecified.

\section{Simulation studies} \setcounter {equation}{0}
\def\theequation{\thesection.\arabic{equation}}
In this section, we investigate the performances of the proposed EDR estimator $\hat{\beta}_{EDR}$ and several other estimators based on Monte-Carlo simulations. For each model and missingness, we generate 1000 Monte Carlo random samples of size $n=200$.

\paragraph{Model 1.}
We consider a scalar response variable $y$ and two-dimensional covariate vector $x=(x_1,x_2)^\textsf{T}$ and models
\begin{eqnarray*}
 y_i=2+3x_{1i}^k+x_{2i}^2+x_{1i}\epsilon_i,\ k=1,2,4,\ i=1,\cdots,n,
\end{eqnarray*}
where $x_{1i}$, $x_{2i}$ and $\epsilon_i$ are independent standard normal random variables. The corresponding estimating function for the mean response is $s(z,\beta)=y-\beta$, where $z=(x^\textsf{T},y)^\textsf{T}$.

We use the missing-data model $\textmd{Logit}\{P(\delta=1|x_{1i},x_{2i}\}= \tau_0+\tau_1x_{1i}+\tau_2x_{2i}+\tau_3x_{1i}x_{2i}$ to generate the non-missing indicator, $\delta_i$, $i=1,\cdots,n$, and the ``working missing-data model" is $\textmd{Logit}\{\pi(x,\gamma)\}= \gamma_0+\gamma_1x_{1}+\gamma_2x_{2}$, where $\textmd{Logit}(u)=\log\{u/(1-u)\}$. Note that when $\tau_3=0$, the missing-data model is specified correctly.

Since $E\{s(z,\beta)|x\}=E(y|x)-\beta$, we set $u(x,\beta,\alpha)=m(x,\alpha)-\beta$, where $m(x,\alpha)$ is a working model for $E(y|x)$. When $\tau_3=0$, we use $m(x,\alpha)=\alpha_0+\alpha_1x_1^2+\alpha_2x_2^2$ and estimate $\alpha=(\alpha_0,\alpha_1,\alpha_2)^\textsf{T}$ by $\hat{\alpha}=\arg\min_{\alpha}\sum_{i=1}^n\delta_i(y_{i}-\alpha_{0}-\alpha_{1}x_{1i}^2-\alpha_{2}x_{2i}^2)^2$. If $k=2$, $m(x,\alpha)$ is correct, whereas $m(x,\alpha)$ is misspecified if $k=1,4$. When $\tau_3\neq0$, we use $m(x,\alpha)=\alpha_0+\alpha_1x_1^k+\alpha_2x_2^2$ as the working model for $E(y|x)$ and estimate $\alpha=(\alpha_0,\alpha_1,\alpha_2)^\textsf{T}$ by $\hat{\alpha}=\arg\min_{\alpha}\sum_{i=1}^n\delta_i(y_{i}-\alpha_{0}-\alpha_{1}x_{1i}^k-\alpha_{2}x_{2i}^2)^2$, $k=1,2,4$.

Seven estimators of $\beta$ are considered. The first one is the sample mean $\hat{\beta}_{ALL}=n^{-1}\sum_{i=1}^ny_i$ with no missing data. This is the ideal case, and we use it as a benchmark for comparison. The second one is the CCA estimator $\hat{\beta}_{CCA}=n^{-1}\sum_{i=1}^n\delta_iy_i$. The third one is Horvitz and Thompsom's estimator $\hat{\beta}_{HT}=n^{-1}\sum_{i=1}^n\delta_iy_i/\pi(x_i,\hat{\gamma})$. The fourth one is the estimator of Robins et al. (1994), $\hat{\beta}_{RRZ}=n^{-1}\sum_{i=1}^n[\delta_iy_i-\{\delta_i-\pi(x_i,\hat{\gamma})\}m(x_i,\hat{\alpha})]/\pi(x_i,\hat{\gamma})$. The fifth one is the Tang and Qin's (2012) estimator, $\hat{\beta}_{TQ}$, which involves the kernel function and the bandwidth. We take the kernel function as $K(u)= \prod _{i=1}^2 K_1(u_i)$, where $K_1(u)=\exp(-u^2/2)/\sqrt{2\pi}$, and set the bandwidth as $h=n^{-\frac{1}{3}}$. The sixth one is the estimator of Qin, Shao and Zhang (2008), $\hat{\beta}_{QSZ}$. The final one is EDR estimator $\hat{\beta}_{EDR}$ defined in (\ref{edrmean}). To compute $\hat{\beta}_{RRZ}$, $\hat{\beta}_{QSZ}$ and $\hat{\beta}_{EDR}$, one needs to impose a parametric model on $E(y|x)$. We use $m(x,\hat{\alpha})$ as a working model for $E(y|x)$ and the settings for $m(x,\hat{\alpha})$ is described as above.

Table 1 shows the empirical bias and the root-mean-squared errors (RMSEs) of the estimators under Model 1 with different missingness. The results in Table 1 can be summarized as follows:

\begin{itemize}
\item If data is MCAR $(\sum_{i=1}^3|\tau_i|=0)$, then all estimators perform well in terms of biases and RMSEs.

\item If data is MAR $(\sum_{i=1}^3|\tau_i|\neq0)$, then the estimators $\hat{\beta}_{CCA}$ and $\hat{\beta}_{TQ}$ for the mean response $\beta$ are clearly biased.

\item When $\pi(x,\gamma)$ and $m(x,\alpha)$ are both specified correctly $(\tau_3=0, k=2)$, biases of  $\hat{\beta}_{HT}$, $\hat{\beta}_{RRZ}$, $\hat{\beta}_{QSZ}$ and $\hat{\beta}_{EDR}$ are negligible. The RMSEs of $\hat{\beta}_{RRZ}$, $\hat{\beta}_{QSZ}$ and $\hat{\beta}_{EDR}$ are almost the same ($\hat{\beta}_{RRZ}$, $\hat{\beta}_{QSZ}$ and $\hat{\beta}_{EDR}$ are asymptotically equivalent in this case). In terms of RMSEs, $\hat{\beta}_{HT}$ is worse than $\hat{\beta}_{RRZ}$, $\hat{\beta}_{QSZ}$ and $\hat{\beta}_{EDR}$, and it can be much worse when the missing rate is high.

\item When $\pi(x,\gamma)$ is specified correctly and $m(x,\alpha)$ is misspecified $(\tau_3=0, k=1,4)$, $\hat{\beta}_{HT}$, $\hat{\beta}_{RRZ}$, $\hat{\beta}_{QSZ}$ and $\hat{\beta}_{EDR}$ are robust. However, the RMSE of $\hat{\beta}_{EDR}$ is smaller than those of $\hat{\beta}_{HT}$, $\hat{\beta}_{RRZ}$ and $\hat{\beta}_{QSZ}$.  In the case of $(\tau_3=0,k=4)$, the RMSE of $\hat{\beta}_{HT}$ is much larger than those of $\hat{\beta}_{RRZ}$, $\hat{\beta}_{QSZ}$ and $\hat{\beta}_{EDR}$.

\item When $\pi(x,\gamma)$ is misspecified and $m(x,\alpha)$ is specified correctly $(\tau_3\neq0, k=1,2,4)$, the Horvitz and Thompsom's estimator $\hat{\beta}_{HT}$ for the mean response $\beta$ is clearly biased. In the case of $(\tau_0,\tau_1,\tau_2,\tau_3)=(0,0.5,1,1)$, the RMSEs of $\hat{\beta}_{RRZ}$, $\hat{\beta}_{QSZ}$ and $\hat{\beta}_{EDR}$ are almost the same. However, in the case of $(\tau_0,\tau_1,\tau_2,\tau_3)=(-1,0.5,1,1)$, the RMSE of $\hat{\beta}_{RRZ}$ is much larger than those of $\hat{\beta}_{QSZ}$ and $\hat{\beta}_{EDR}$.
\end{itemize}

\paragraph{Model 2.}
We consider a two-dimensional response variable $y=(y_1,y_2)^\textsf{T}$ and a scalar covariate $x$ and models
\begin{eqnarray*}
&&y_{1i}=2+3x_i^k+\epsilon_{1i},\\
&&y_{2i}=2+3x_i^k+x_i\epsilon_{2i},\ k=1,2,4,\ i=1,\cdots,n,
\end{eqnarray*}
where $\epsilon_{1i}$, $\epsilon_{2i}$ and $x_i$ are independent standard normal random variables. The corresponding estimating function for the mean response of $y=(y_1,y_2)^\textsf{T}$ is $s(z,\beta)=(y_1-\beta,y_2-\beta)^\textsf{T}$, where $z=(x,y^\textsf{T})^\textsf{T}$.

For each model, we use the missing-data model $\textmd{Logit}\{P(\delta=1|x_{i}\}= \tau_0+\tau_1x_i+\tau_2x_i^2 $ to generate the non-missing indicator, $\delta_i$, $i=1,\cdots,n$, and the ``working missing-data model" is $\textmd{Logit}\{\pi(x,\gamma)\}= \gamma_0+\gamma_1x$. Note that when $\tau_2=0$, the missing-data model is specified correctly.

Since $E\{s(z,\beta)|x\}=(E(y_1|x)-\beta,E(y_2|x)-\beta)^\textsf{T}$, we set $u(x,\beta,\alpha)=(m_1(x,\alpha)-\beta,m_2(x,\alpha)-\beta)^\textsf{T}$, where $m_j(x,\alpha)$ is a working model for $E(y_j|x)$, $j=1,2$. When $\tau_2=0$, we use $m_j(x,\alpha)=\alpha_{j0}+\alpha_{j1}x^2$ as the working model for $E(y_j|x)$, where $\alpha=(\alpha_{10},\alpha_{11},\alpha_{20},\alpha_{21})^\textsf{T}$. We estimate $\alpha$ by $\hat{\alpha}=\arg\min_{\alpha}\sum_{i=1}^n\delta_i\sum_{j=1}^2(y_{ji}-\alpha_{j0}-\alpha_{j1}x_i^2)^2$. If $k=2$, $u(x,\beta,\alpha)$ is correct, whereas $u(x,\beta,\alpha)$ is misspecified if $k=1,4$. When $\tau_2\neq0$, we use $m_j(x,\alpha)=\alpha_{j0}+\alpha_{j1}x^k$ as the working model for $E(y_j|x)$, $k=1,2,4$, and estimate $\alpha$ by $\hat{\alpha}=\arg\min_{\alpha}\sum_{i=1}^n\delta_i\sum_{j=1}^2(y_{ji}-\alpha_{j0}-\alpha_{j1}x_i^k)^2$.

Six estimators of $\beta$ are considered. The first one is the EL estimator (Qin, 1994) using the estimating function $s(z,\beta)=(y_1-\beta,y_2-\beta)^\textsf{T}$ with no missing data. The second one is the CCA estimator $\hat{\beta}_{CCA}$, which is the EL estimator using the estimating function $s(z,\beta)=(y_1-\beta,y_2-\beta)^\textsf{T}$ with complete-case data. The third one is Horvitz and Thompsom's estimator $\hat{\beta}_{HT}$. The fourth one is the RRZ estimator, $\hat{\beta}_{RRZ}$. The fifth one is the Tang and Qin's (2012) estimator, $\hat{\beta}_{TQ}$, where the kernel function is $K_1(u)$ and the bandwidth is $h=n^{-\frac{1}{3}}$. The final one is EDR estimator $\hat{\beta}_{EDR}$. To compute $\hat{\beta}_{RRZ}$ and $\hat{\beta}_{EDR}$, we use $m_j(x,\hat{\alpha})$ as a working model for $E(y_j|x)$ and the settings for $m_j(x,\hat{\alpha})$ is described as above.

Table 2 shows the empirical bias and the root-mean-squared errors (RMSEs) of the proposed estimators under Model 2 with different missingness. The results in Table 2 can be summarized as follows:
\begin{itemize}
\item If data is MCAR $(\tau_1=\tau_2=0)$, then all estimators perform well in terms of biases and RMSEs.

\item If data is MAR $(\tau_1\neq0\ \mbox{or}\ \tau_2\neq0)$, then the estimators $\hat{\beta}_{CCA}$ and $\hat{\beta}_{TQ}$ for the mean response $\beta$ are clearly biased. However, $\hat{\beta}_{TQ}$ is robust in terms
      of the RMSEs.

\item When $\pi(x,\gamma)$ and $u(x,\beta,\alpha)$ are both specified correctly $(\tau_2=0, k=2)$, biases of  $\hat{\beta}_{HT}$, $\hat{\beta}_{RRZ}$, and $\hat{\beta}_{EDR}$ are negligible. The RMSEs of $\hat{\beta}_{RRZ}$ and $\hat{\beta}_{EDR}$ are almost the same ($\hat{\beta}_{RRZ}$ and $\hat{\beta}_{EDR}$ are asymptotically equivalent in this case). In terms of RMSEs, $\hat{\beta}_{HT}$ is worse than $\hat{\beta}_{RRZ}$ and $\hat{\beta}_{EDR}$, and it can be much worse when the missing rate is high.

\item When $\pi(x,\gamma)$ is specified correctly and $u(x,\beta,\alpha)$ is misspecified $(\tau_2=0, k=1,4)$, $\hat{\beta}_{HT}$, $\hat{\beta}_{RRZ}$, $\hat{\beta}_{QSZ}$ and $\hat{\beta}_{EDR}$ are robust. In the case of $(\tau_2=0,k=1)$, the RMSEs of $\hat{\beta}_{HT}$ and $\hat{\beta}_{EDR}$ are almost the same and are smaller than those of  $\hat{\beta}_{RRZ}$. In the case of $(\tau_2=0,k=4)$, $\hat{\beta}_{HT}$ has the largest RMSE and $\hat{\beta}_{EDR}$ has the smallest RMSE.

\item When $\pi(x,\gamma)$ is misspecified and $u(x,\beta,\alpha)$ is specified correctly $(\tau_3\neq0, k=1,2,4)$, the Horvitz and Thompsom's estimator $\hat{\beta}_{HT}$ for the mean response $\beta$ is clearly biased. The RMSEs of $\hat{\beta}_{TQ}$, $\hat{\beta}_{EDR}$, $\hat{\beta}_{RRZ}$ are almost the same and much smaller than that of $\hat{\beta}_{HT}$.
\end{itemize}

\paragraph{Model 3.}

We consider a scalar response variable $x_2$ and two-dimensional covariate vector $(x_1, y)^\textsf{T}$ and models
\begin{eqnarray*}
x_{2i}=\beta_0+\beta_1x_{1i}+\beta_2y_i+\epsilon_i,\ i=1,\cdots,n,
\end{eqnarray*}
where $\epsilon_i$ and $(x_{1i},y_i)$ are independent, $\epsilon_i\sim N (0,1)$, $x_{1i}\sim Exp(1)$ and $y_i\sim N \chi^(1)$. The corresponding estimating function for the regression coefficient $\beta=(1,1,1)$ is $s(z,\beta)= (1,x_1,y)^\textsf{T}\{ x_2- (\beta_0+\beta_1x_1+\beta_2y)\}$, where $\beta=(\beta_0,\beta_1,\beta_2)^\textsf{T}$, $z=(x^\textsf{T},y)^\textsf{T}$ and $x=(x_1,x_2)^\textsf{T}$.

For each model, we use the missing-data model $\textmd{Logit}\{P(\delta=1|x_{1i},x_{2i}\}= \tau_0+\tau_1x_{1i}+\tau_2x_{2i}+\tau_3 x_{1i}x_{2i} $ to generate the non-missing indicator, $\delta_i$, $i=1,\cdots,n$, and the ``working missing-data model" is $\textmd{Logit}\{\pi(x,\gamma)\}= \gamma_0+\gamma_1x_{1}+\gamma_2x_{2}+\gamma_3 x_{1}x_{2}$. Thus, the missing-data model is specified correctly.

Seven estimators of $\beta$ are considered. The first one is the OLS estimator $\hat{\beta}_{ALL}=\arg\min_{\beta}\sum_{i=1}^n\{x_{2i}- (\beta_0+\beta_1x_{1i}+\beta_2y_i)\}^2$ with no missing data. The second one is the CCA estimator $\hat{\beta}_{CCA}=\arg\min_{\beta}\sum_{i=1}^n\delta_i\{x_{2i}- (\beta_0+\beta_1x_{1i}+\beta_2y_i)\}^2$. The third one is Horvitz and Thompsom's estimator $\hat{\beta}_{HT}=\arg\min_{\beta}\sum_{i=1}^n\delta_i\{x_{2i}- (\beta_0+\beta_1x_{1i}+\beta_2y_i)\}^2/\pi(x_i,\hat{\gamma})$. The fourth one is the RRZ estimator $\hat{\beta}_{RRZ}$. The fifth one is the Tang and Qin's (2012) estimator, $\hat{\beta}_{TQ}$, where the kernel function is $K(u)$ and the bandwidth is $h=n^{-\frac{1}{3}}$. The sixth one is the estimator of Qin, Zhang and Leung (2009), $\hat{\beta}_{QZL}$. The final one is EDR estimator $\hat{\beta}_{EDR}$. To compute $\hat{\beta}_{RRZ}$, $\hat{\beta}_{QZL}$ and $\hat{\beta}_{EDR}$, one needs to impose a parametric model on $E\{s(z,\beta)|x\}$. We use $u(z,\beta,\alpha)= (1,x_1,\hat{y}(\alpha))^\textsf{T}\{ x_2- (\beta_0+\beta_1x_{1}+\beta_2\hat{y}(\alpha))\}$ as the working model for $E\{s(z,\beta)|x\}$, where $\hat{y}(\alpha)=\alpha_0+ \alpha_1x_{1}+\alpha_2x_{2}$ and $\alpha=(\alpha_0,\alpha_1,\alpha_2)^\textsf{T}$. We estimate $\alpha$ by $\hat{\alpha}=\arg\min_{\alpha}\sum_{i=1}^n\delta_i(y_i-\alpha_0- \alpha_1x_{1i}-\alpha_2x_{2i})^2$.

Table 3 shows the empirical bias and the root-mean-squared errors (RMSEs) of the proposed estimators under Model 3 with different missingness. The results in Table 3 can be summarized as follows:
\begin{itemize}

\item If data is MCAR $(\tau_1=\tau_2=0)$, then all estimators perform well in terms of biases and RMSEs.

\item If data is MAR $(\tau_1\neq0\ \mbox{or}\ \tau_2\neq0)$, then the estimator $\hat{\beta}_{CCA}$ for the regression coefficient $\beta$ is clearly biased. Moreover, biases of  $\hat{\beta}_{HT}$, $\hat{\beta}_{RRZ}$, $\hat{\beta}_{QZL}$ and $\hat{\beta}_{EDR}$ are negligible.

\item In the cases of $(\tau_0,\tau_1,\tau_2)=(1,1,1)$, $(0,1,1)$ and $(1,-1,1)$, $\hat{\beta}_{TQ}$ has the smallest RMSE and the RMSEs of $\hat{\beta}_{TQ}$, $\hat{\beta}_{QZL}$ and $\hat{\beta}_{EDR}$ are smaller than those of $\hat{\beta}_{HT}$ and $\hat{\beta}_{RRZ}$.

\item In the cases of $(\tau_0,\tau_1,\tau_2)=(-1,1,-1)$, the RMSE of $\hat{\beta}_{HT}$ is much larger than those of $\hat{\beta}_{TQ}$, $\hat{\beta}_{QZL}$ and $\hat{\beta}_{EDR}$.

\end{itemize}

Based on these simulation results, we can draw the following conclusions:

\begin{enumerate}

\item The Horvitz-Thompson estimator $\hat{\beta}_{HT}$, which
is robust against the misspecification of regression model $u(z,\beta,\alpha)$ but not robust against the misspecification of propensity model $\pi(x,\gamma)$, and is an inefficient estimator.

\item The Robins-Rotnitzky-Zhao estimator $\hat{\beta}_{RRZ}$ and our proposed estimator $\hat{\beta}_{EDR}$ are doubly robust, but $\hat{\beta}_{EDR}$ is more
efficient than $\hat{\beta}_{RRZ}$ when the working model $u(z,\beta,\alpha)$ is misspecified, regardless of whether or not the propensity model $\pi(x,\gamma)$ is
specified correctly.

\item The performance of Tang and Qin's (2012) estimator, $\hat{\beta}_{TQ}$, depends on the choice of bandwidth.
\end{enumerate}

Table 4-6 give the empirical variances, the mean of estimated variances of the estimators $\hat{\beta}_{HT}$, $\hat{\beta}_{RRZ}$ and $\hat{\beta}_{EDR}$, and the empirical coverage probabilities of 95\%  confidence intervals of $\beta$ for Models 1-3, respectively.  The estimated variance of $\hat{\beta}_{HT}$ and corresponding confidence interval of $\beta$ perform well only when $\pi(x,\gamma)$ is specified correctly and $k=1,2$. The estimated variance of $\hat{\beta}_{RRZ}$ and $\hat{\beta}_{EDR}$ and corresponding confidence interval of $\beta$ perform well in all cases, except for the case $k=4$, in which the sample size $n = 200$ may not be large enough to show the asymptotic effect.

\section{Data analysis}
\setcounter {equation}{0}
\def\theequation{\thesection.\arabic{equation}}

In this section, we apply the proposed method to an economics data, which were collected by Lalonde (1986). Dehejia and Wahba (1999) used propensity score methods to estimate the treatment effect of a labour training programme called `National support work demonstration' on postintervention earnings. Here
we use a subset of the data that were used by Lalonde (1986), Dehejia and Wahba (1999), Abadie \emph{et al.} (2004), and Qin \emph{et al.} (2008). The data set contains 445 individuals. There are 185 individuals participated in the training program and 260 individuals did not participate. The possible covariates are age, years of education, an indicator for African-American, an indicator for Hispanic-American, marital status, an indicator for more than grade school but less than high school education and earnings in 1974 and 1975.

Let $\mu_1$ and $\mu_0$ be the population mean of earnings in 1978 for individuals participating and not participating in the training program, respectively. We are interested in estimating $\Delta=\mu_1-\mu_0$, which is the potential effect of participation in this job training programme on individuals' earnings in 1978. To estimate $\Delta$, we only need to estimate $\mu_1$ and $\mu_0$, respectively. In the estimation of $\mu_1$, we treat the observations from 185 individuals participating in the training program as completely observed data and treat the rest of 260 individuals as missing data (i.e., their treatment responses as participated individuals are considered missing). Meanwhile, we treat the observations from 260 individuals not participating in the training program as completely observed data and treat the rest of 185 individuals as missing data in the estimation of $\mu_0$.

To apply our method, our first step is to consider a working model for the propensity function. We used a logistic propensity model and chose the covariates by the stepwise search algorithm ``stepAIC" in R package $\verb"MASS"$. Two variables were included in the logistic propensity score analysis: an indicator for Hispanic-American and an indicator for more than grade school but less than high school education. Our second step is to pick working regression models. We examined the regression model with possible covariates and earnings in 1978 separately in the two groups of individuals, and chose the covariates by the stepwise search algorithm ``stepAIC" in R package $\verb"MASS"$. For individuals participating in the training program, we chose a working linear regression model with one covariate: education. For individuals not participating in the training program, we chose a linear regression model with two covariates:  an indicator for African-American, and earnings in 1974. We applied the proposed procedure separately to two groups of individuals and obtained estimates of $\mu_1$ and $\mu_0$. Then, $\Delta$ is estimated by the difference between the estimated $\mu_1$ and $\mu_0$. For comparison, we also applied the other methods considered in Model 1 of Section 5 with the same working propensity and regression models.

Point estimates, bootstrap standard errors and the asymptotic variance formula-based standard errors are reported in Table 7. To calculate bootstrap standard errors of all estimators, the bootstrap replications is set to be 500.  The asymptotic variance formula-based standard errors are only reported for estimators $\hat{\beta}_{HT}$, $\hat{\beta}_{RRZ}$ and $\hat{\beta}_{EDR}$. From Table 7, we can see that all estimates demonstrate at least a \$1,600 increase from participating in the training program. Our proposed estimate $\hat{\beta}_{EDR}$ is nearly the same as the Robins-Rotnitzky-Zhao estimate $\hat{\beta}_{RRZ}$ and Qin-Shao-Zhang estimate $\hat{\beta}_{QSZ}$, indicating that our
working models are reasonable. The other three estimates, $\hat{\beta}_{CCA}$, $\hat{\beta}_{TQ}$ and $\hat{\beta}_{HT}$ are somewhat different than $\hat{\beta}_{EDR}$. The CCA estimate for $\mu_0$ and $\mu_1$ is larger than any other estimates, indicating a possible positive bias in the estimation of $\mu_0$ and $\mu_1$. Moreover, the asymptotic variance formula-based standard error is nearly the same as bootstrap standard error for estimators $\hat{\beta}_{HT}$, $\hat{\beta}_{RRZ}$ and $\hat{\beta}_{EDR}$.

\section{Conclusion} \setcounter {equation}{0}
\def\theequation{\thesection.\arabic{equation}}
In this paper, we propose an EDR approach for making inference about the parameter vector defined by EEs when data are missing at random. As with the semiparametric estimation procedure that was proposed by Robins et al. (1994), the EDR inference procedure also enjoys the double-robustness property, i.e. the EDR estimator is consistent when either the propensity model or regression model for the conditional expectation $E\{s(x,y,\beta)|x\}$ is correctly specified. We established some asymptotic results. In particular, the proposed EDR estimators can achieve the semiparametric efficiency bound in the sense of Bickel et al. (1993) if the propensity model is specified correctly. Moreover, if both the propensity model and regression model for the conditional expectation $E\{s(x,y,\beta)|x\}$ are specified correctly, the EDR estimator can achieve semiparametric efficiency lower bound in the sense of Chen et al. (2008). In addition, we developed the asymptotic covariance formula-based doubly robust estimator of the asymptotic covariance of the EDR estimator. Thus, statistical inference based on our approach requires neither resampling nor kernel smoothing. Simulation results show that the proposed estimator is competitive against existing estimation method.

One may include a nonparametric estimator of $E\{s(z,\beta)|x\}$ in $a(x,\beta,\alpha)$ in (\ref{xi}), to obtain a more robust estimator $\hat{\beta}_{EDR}$. But such additional robustness is complicated by the sensitivity of bandwidth selection and a possible loss of efficiency (Remark \ref{afun}), especially when $x$ is of relatively  high dimension.

\section*{Acknowledgements}
Tianqing Liu was partly supported by the NSFC (No.11201174) and the Natural Science Foundation for Young Scientists
of Jilin Province, China (No.20150520054JH); Xiaohui Yuan was partly supported by the NSFC (No. 11401048) and the Natural Science Foundation for Young Scientists of Jilin
Province, China (No.20150520055JH).

\appendix
\section*{Appendix}
\newcounter{lemm}
\newtheorem{lemm}{Lemma A.}
\setcounter {equation}{0}
\def\theequation{A.\arabic{equation}}

Unless mentioned otherwise, all limits are taken as $n\rightarrow\infty$ and $\|\cdot\|$ denotes the Euclidean norm. For notational convenience, for $i=1\cdots,n$, let $U_{1i}=U_{1}(t_i,\gamma^*)$, $h_i=h(t_i,\beta^*,\gamma^*,\alpha^*)$, $g_i=g(t_i,\beta^*,\gamma^*,\alpha^*)$, $\varphi_{1i}=\varphi_1(t_i,\beta^*,\gamma^*)$ and
$G_\gamma=E\left\{\frac{\partial g(t_i,\beta^*,\gamma^*,\alpha^*)}{\partial \gamma^\textsf{T}}\right\}$. To establish the large sample properties in this paper, we require the following conditions:
\vskip0.2cm \noindent\textbf{Regularity Conditions}
\\
C1: $\{t_{i}=(z_i^\textsf{T},\delta_i)^\textsf{T}\}_{i=1}^n$ are independent and identically distributed.
\\
C2: Let $\theta^*=(\gamma^{*\textsf{T}},\alpha^{*\textsf{T}},\lambda^{*\textsf{T}})^\textsf{T}$ and $\vartheta^*=(\beta^{*\textsf{T}},\theta^{*\textsf{T}})^\textsf{T}$. We require that $\vartheta^*$ be an interior point of a compact parameter space $\Psi \subset \mathcal {R}^d$, where $d$ is the dimension of $\vartheta^*$.
\\
C3: $\omega(x)\cdot\pi(x,\gamma^*)$ is bounded away from zero, i.e. $\inf_{x}\{\omega(x)\cdot\pi(x,\gamma^*)\}\geq{c_0}$ for some $c_0>0$.
\\
C4: $\varphi_{3}(t_i, \vartheta)$ and $U(t_i, \vartheta)$ are continuously differentiable at each $\vartheta\in\Psi$ with probability one. $E\{\sup_{\vartheta\in\Psi}\|\varphi_{3}(t_i, \vartheta)\|^2\}<\infty$, $E\{\sup_{\vartheta\in\Psi}\|U(t_i, \vartheta)\|^2\}<\infty$, $E\{\sup_{\vartheta\in\Psi}\|\partial\varphi_{3}(t_i, \vartheta)/\partial\vartheta^\textsf{T}\|\}<\infty$, $E\{\sup_{\vartheta\in\Psi}\|\partial U(t_i, \vartheta)/\partial\vartheta^\textsf{T}\|\}<\infty$; $V_{31}(\vartheta)$ and $E\{\partial U(t_i, \vartheta)/\partial\theta^\textsf{T}\}$ are nonsingular for $\vartheta\in\Psi$.
\\
C5: Let $V_3(\vartheta)=V_{32}^\textsf{T}(\vartheta)V_{31}^{-1}(\vartheta)$, $\eta_{3}(t_i,\vartheta)=((V_3(\vartheta)\varphi_{3}(t_i,\vartheta))^\textsf{T},U^\textsf{T}(t_i,\vartheta))^\textsf{T}$, and $\eta_{30}(\vartheta)=E\{\eta_3(t_i,\vartheta)\}$. For $\vartheta\in\Psi$, $\eta_{30}(\vartheta)=0$ only if $\vartheta=\vartheta^*$.
\\
C6: $\left(\begin{array}{c}
V_3(\vartheta^*)E\{\partial\varphi_{3}(t_i,\vartheta^*)/\partial\vartheta^\textsf{T}\}\\
E\{\partial U(t_i,\vartheta^*)/\partial\vartheta^\textsf{T}\}\\
\end{array}\right)$ and $V_{32}^\textsf{T}(\vartheta^*)V_{31}^{-1}(\vartheta^*)V_{32}(\vartheta^*)$ are nonsingular.

\noindent{\bf Proof of Theorem \ref{HT}} The proof is similar to that of Theorem \ref{elw0} and the details are omitted.

\noindent{\bf Proof of Theorem \ref{RRZ}} The proof is similar to that of Theorem \ref{elw0} and the details are omitted.

\begin{lemm} \label{identification}
If  $\pi(x,\gamma^*)=\omega(x)$ or $u(x,\beta,\alpha^*)=E\{s(z,\beta)|x\}$, then $E\{\varphi_{3}(t_i,\vartheta^*)\}=0$ and $\eta_{30}(\vartheta^*)=0$.
\end{lemm}
\noindent{\bf Proof of Lemama A.\ref{identification}} If $\pi(x,\gamma^*)=\omega(x)$, it is easy to verify that $\lambda^*=\lambda(\beta^*,\gamma^*,\alpha^*)=0$. Thus, $E\{\varphi_{3}(t_i,\vartheta^*)\}=E\{\frac{\delta_i g(z_i,\beta^*)}{\pi(x_i,\gamma^*)}\}=E\{g(z_i,\beta^*)\}=0$. If $u(x,\beta,\alpha^*)=E\{s(z,\beta)|x\}$, we have
\begin{eqnarray*}
&&E\{\varphi_{3}(t_i,\vartheta^*)\}\\
&=&E\left\{\frac{1}{1+\lambda^{*\textsf{T}}g(t_i,\beta^*,\gamma^*,\alpha^*)}\frac{\delta_i \{s(z_i,\beta^*)-u(x_i,\beta^*,\alpha^*)\}}{\pi(x_i,\gamma^*)}\right\}\\
&&+E\{u(x_i,\beta^*,\alpha^*)\}\\
&=&E\left[\frac{\delta_i }{\pi(x_i,\gamma^*)\{1+\lambda^{*\textsf{T}}g(t_i,\beta^*,\gamma^*,\alpha^*)\}}E\{s(z_i,\beta^*)-u(x_i,\beta^*,\alpha^*)|x_i\}\right]\\
&&+E\{s(z_i,\beta^*)\}\\
&=&0.
\end{eqnarray*}
\begin{lemm} \label{consistent}
Suppose that the regularity conditions C1-C5 hold. Let $\hat{\vartheta}=(\hat{\beta}_{EDR}^\textsf{T},\hat{\theta}^\textsf{T})^\textsf{T}$ with $\hat{\theta}=\hat{\theta}(\hat{\beta}_{EDR})=(\hat{\gamma}^\textsf{T},\hat{\alpha}^\textsf{T},\hat{\lambda}^\textsf{T}(\hat{\beta}_{EDR},\hat{\gamma},\hat{\alpha}))^\textsf{T}$.
Then, $\hat{\vartheta}\prow\vartheta^*$ as $n\rightarrow\infty$.
\end{lemm}
\noindent{\bf Proof of Lemama A.\ref{consistent}}. Let $\hat{\eta}_{3}(t_i,\vartheta)=((\hat{V}_3(\vartheta)\varphi_{3}(t_i,\vartheta))^\textsf{T},U^\textsf{T}(t_i,\vartheta))^\textsf{T}$ with $\vartheta=(\beta^\textsf{T},\theta^\textsf{T})^\textsf{T}$. Then, $\hat{\vartheta}$ can be written as
\begin{eqnarray*}
\hat{\vartheta}&=&\arg\max_{\vartheta}\left\{-\frac{1}{2}\left\|n^{-1}\sum_{i=1}^n\hat{\eta}_3(t_i,\vartheta)\right\|^{2}\right\}.
\end{eqnarray*}
From condition C4 and Lemma 2.4 in Newey and McFadden (1994), it follow that $\sup_{\vartheta\in\Psi}\|\hat{C}_{1}(\vartheta)-C_{1}(\vartheta)\|=o_p(1)$ and $\sup_{\vartheta\in\Psi}\|\hat{C}_{2}(\vartheta)-C_{2}(\vartheta)\|=o_p(1)$, where
\begin{eqnarray*}
&&\hat{C}_{1}(\vartheta)=n^{-1}\sum_{i=1}^n\partial \varphi_{3}(t_i, \vartheta)/\partial  \theta^\textsf{T},\ \ \hat{C}_2(\vartheta)=n^{-1}\sum_{i=1}^n\partial U(t_i,\vartheta)/ \partial\theta^\textsf{T},\\
&&{C}_{1}(\vartheta)=E\{\varphi_{3}(t_i, \vartheta)/\partial  \theta^\textsf{T}\},\ \ C_2(\vartheta)=E\{\partial U(t_i,\vartheta)/ \partial\theta^\textsf{T}\}.
\end{eqnarray*}
Then, $\sup_{\vartheta\in\Psi}\|\hat{C}_{1}(\vartheta)\hat{C}_2^{-1}(\vartheta)-C_{1}(\vartheta)C_2^{-1}(\vartheta)\|=o_p(1)$. Similarly, we can show that $\sup_{\vartheta\in\Psi}\|\hat{V}_{3}(\vartheta)-V_{3}(\vartheta)\|=o_p(1)$ and $V_{3}(\vartheta)$ is continuous on $\Psi$. Based on this fact, it follows that $\sup_{\vartheta\in\Psi}\|n^{-1}\sum_{i=1}^n\hat{\eta}_{3}(t_i,\vartheta)-\eta_{30}(\vartheta)\|=o_p(1)$. Therefore,
$-\frac{1}{2}\left\|n^{-1}\sum_{i=1}^n\hat{\eta}_3(t_i,\vartheta)\right\|^{2}$ converges uniformly in probability to $-\frac{1}{2}\|\eta_{30}(\vartheta)\|^2$. By condition C5 and Lemma A.\ref{identification}, $-\frac{1}{2}\|\eta_{30}(\vartheta)\|^2$ is uniquely maximized at $\vartheta^*$. Using Theorem 2.1 in Newey and McFadden (1994), we have $\hat{\vartheta}\prow\vartheta^*$ as $n\rightarrow\infty$.

\noindent{\bf Proof of Theorem \ref{elw0}} By Lemma A.\ref{consistent} and the mean value theorem, we get
\begin{eqnarray*}
&&0=n^{-1}\sum_{i=1}^n\hat{\eta}_3(t_i,\hat{\vartheta})=\left(\begin{array}{c}
\hat{V}_3(\hat{\vartheta})n^{-1}\sum_{i=1}^n\varphi_{3}(t_i,\hat{\vartheta})\\
n^{-1}\sum_{i=1}^n U(t_i,\hat{\vartheta})\\
\end{array}\right)\\
&&=\left(\begin{array}{c}
\hat{V}_3(\hat{\vartheta})n^{-1}\sum_{i=1}^n\varphi_{3}(t_i,\vartheta^*)\\
n^{-1}\sum_{i=1}^n U(t_i,\vartheta^*)\\
\end{array}\right)+\left(\begin{array}{c}
\hat{V}_3(\hat{\vartheta})n^{-1}\sum_{i=1}^n\{\partial\varphi_{3}(t_i,\bar{\vartheta})/\partial\vartheta^\textsf{T}\}\\
n^{-1}\sum_{i=1}^n\{\partial U(t_i,\bar{\vartheta})/\partial\vartheta^\textsf{T}\}\\
\end{array}\right)(\hat{\vartheta}-\vartheta^*).
\end{eqnarray*}
The above asymptotic expansion yields the following asymptotic expression for $\hat{\vartheta}$:
\begin{eqnarray*}
&&n^{1/2}(\hat{\vartheta}-\vartheta^*)\\
&=&-\left(\begin{array}{c}
\hat{V}_3(\hat{\vartheta})n^{-1}\sum_{i=1}^n\{\partial\varphi_{3}(t_i,\bar{\vartheta})/\partial\vartheta^\textsf{T}\}\\
n^{-1}\sum_{i=1}^n\{\partial U(t_i,\bar{\vartheta})/\partial\vartheta^\textsf{T}\}\\
\end{array}\right)^{-1}\left(\begin{array}{c}
\hat{V}_3(\hat{\vartheta})n^{-1/2}\sum_{i=1}^n\varphi_{3}(t_i,\vartheta^*)\\
n^{-1/2}\sum_{i=1}^n U(t_i,\vartheta^*)\\
\end{array}\right)\\
&=&-\left(\begin{array}{c}
V_3(\vartheta^*)E\{\partial\varphi_{3}(t_i,\vartheta^*)/\partial\vartheta^\textsf{T}\}\\
E\{\partial U(t_i,\vartheta^*)/\partial\vartheta^\textsf{T}\}\\
\end{array}\right)^{-1}\left(\begin{array}{c}
V_3(\vartheta^*)n^{-1/2}\sum_{i=1}^n\varphi_{3}(t_i,\vartheta^*)\\
n^{-1/2}\sum_{i=1}^n U(t_i,\vartheta^*)\\
\end{array}\right)+o_p(1)\\
&=&-K_1^{-1}n^{-1/2}\sum_{i=1}^n\eta_3(t_i,\vartheta^*)+o_p(1),
\end{eqnarray*}
where $\bar{\vartheta}$ is a point on the segment connecting $\hat{\vartheta}$ and $\vartheta^*$, and
\begin{eqnarray*}
K_1&=&\left(\begin{array}{c}
V_3(\vartheta^*)E\{\partial\varphi_{3}(t_i,\vartheta^*)/\partial\vartheta^\textsf{T}\}\\
E\{\partial U(t_i,\vartheta^*)/\partial\vartheta^\textsf{T}\}
\end{array}\right)\\
&=&\left(\begin{array}{cc}
V_3(\vartheta^*)E\{\frac{\partial\varphi_{3}(t_i,\vartheta^*)}{\partial\beta^\textsf{T}}\}&V_3(\vartheta^*)E\{\frac{\partial\varphi_{3}(t_i,\vartheta^*)}{\partial\theta^\textsf{T}}\}\\
E\{\frac{\partial U(t_i,\vartheta^*)}{\partial\beta^\textsf{T}}\}&E\{\frac{\partial U(t_i,\vartheta^*)}{\partial\theta^\textsf{T}}\}
\end{array}\right)\\
&=&\left(\begin{array}{cc}
K_{11}&K_{12}\\
K_{21}&K_{22}
\end{array}\right).
\end{eqnarray*}
By the inverse formula for $2 \times2$ block matrices and the multivariate central limit theorem, we have
\begin{eqnarray*}
&&n^{1/2}(\hat{\beta}_{EDR}-\beta^*)\\
&=&-K_{11.2}^{-1}n^{-1/2}\sum_{i=1}^n\{V_3(\vartheta^*)\varphi_{3}(t_i,\vartheta^*)-K_{12}K_{22}^{-1}U(t_i,\vartheta^*)\}+o_p(1)\\
&\drow&N(0,K_{11.2}^{-1}),\ \ \mbox{as}\ \ n\rightarrow\infty,
\end{eqnarray*}
where $K_{11.2}=K_{11}-K_{12}K_{22}^{-1}K_{21}=V_3(\vartheta^*)V_{32}(\vartheta^*)=V_{32}^\textsf{T}(\vartheta^*)V_{31}^{-1}(\vartheta^*)V_{32}(\vartheta^*)$.

\noindent{\bf Proof of Theorem \ref{influence}}
If $\pi(x,\gamma^*)=\omega(x)$, recall that $\lambda^*=0$. Moreover, it is easy to verify that $E\left(\frac{\partial{U(t_i,\vartheta^*)}}{\partial\beta^\textsf{T}}\right)=0$. Based on this fact, we have $V_{32}(\vartheta^*)=E\left\{\frac{\partial\varphi_3(t_i, \vartheta^*)}{\partial\beta^\textsf{T}}\right\}=F_{\beta}^\textsf{T}$.
We can write
\begin{eqnarray*}
&&E\left(\frac{\partial \varphi_{3}}{\partial  \theta^\textsf{T}} \right)=\left(F_{\gamma}\vdots0\vdots -F_{g}\right).
\end{eqnarray*}
and
\begin{eqnarray*}
E\left(\frac{\partial{U}}{\partial\theta^\textsf{T}}\right)&=&\left(\begin{array}{ccc}
-S_B&0&0\\
0&E\left(\frac{\partial{U_2}}{\partial\alpha^\textsf{T}}\right)&0\\
G_{\gamma}&0&-S_g\\
\end{array}\right).
\end{eqnarray*}
where $S_B=E(U_1U_1^\textsf{T})$. Then, we can use the expression of $E\left(\frac{\partial{U(t_i,\vartheta^*)}}{\partial\theta^\textsf{T}}\right)$ and some straightforward algebra to show that
\begin{eqnarray*}
\left\{E\left(\frac{\partial{U(t_i,\vartheta^*)}}{\partial\theta^\textsf{T}}\right)\right\}^{-1}&=&\left(\begin{array}{ccc}
-S_B^{-1}&0&0\\
0&\{E\left(\frac{\partial{U_2}}{\partial\alpha^\textsf{T}}\right)\}^{-1}&0\\
-S_g^{-1}G_{\gamma}S_B^{-1}&0&-S_g^{-1}\\
\end{array}\right).
\end{eqnarray*}
Next, we show that
\begin{eqnarray}\label{ifc}
&&\varphi_{3}(t_i, \vartheta^*)-E\left\{\frac{\partial \varphi_{3}(t_i, \vartheta^*)}{\partial  \theta^\textsf{T}} \right\} \left(E\left\{\frac{\partial U(t_i,\vartheta^*)}{\partial  \theta^\textsf{T}}\right\}\right)^{-1}U(t_i,\vartheta^*)\nonumber\\
&=&\varphi_{1i}-F_gS_g^{-1}g_i.
\end{eqnarray}
Since $\varphi_{3}(t_i, \vartheta^*)=\varphi_{1i}$, we only need to show
\begin{eqnarray}\label{v31}
E\left\{\frac{\partial \varphi_{3}(t_i, \vartheta^*)}{\partial  \theta^\textsf{T}} \right\} \left(E\left\{\frac{\partial U(t_i,\vartheta^*)}{\partial  \theta^\textsf{T}}\right\}\right)^{-1}U(t_i,\vartheta^*)=F_gS_g^{-1}g_i.
\end{eqnarray}
Moreover, applying the inverse formula for $2 \times2$ block matrices and using $F_{\gamma}=E\left(\frac{\partial \varphi_{1}}{\partial  \gamma^\textsf{T}} \right)=-E(\varphi_{1}U_1^\textsf{T})$ and $H_{\gamma}=E(\frac{\partial h}{\partial\gamma^\textsf{T}})=-E(hU_1^\textsf{T})$, we obtain
\begin{eqnarray}\label{identity}
F_{\gamma}-F_{g}S_{g}^{-1}G_\gamma=0.
\end{eqnarray}
Utilizing the identity (\ref{identity}), it follows that
\begin{eqnarray*}
&&E\left\{\frac{\partial \varphi_{3}}{\partial  \theta^\textsf{T}} \right\} \left(E\left\{\frac{\partial U}{\partial  \theta^\textsf{T}}\right\}\right)^{-1}U(t_i,\vartheta^*)\\
&=&\{-F_{\gamma}+F_{g}S_{g}^{-1}G_\gamma\}S_B^{-1}U_{1i}+F_gS_g^{-1}g_i\\
&=&F_gS_g^{-1}g_i\\
&=&(F_h-F_\gamma
S_B^{-1}H_{\gamma}^\textsf{T})(S_h-H_{\gamma}S_B^{-1}H_{\gamma}^\textsf{T})^{-1}h_i\nonumber\\
&&-\{F_\gamma-(F_h-F_\gamma
S_B^{-1}H_{\gamma}^\textsf{T})(S_h-H_{\gamma}S_B^{-1}H_{\gamma}^\textsf{T})^{-1}H_{\gamma}\}S_B^{-1}U_{1i}.
\end{eqnarray*}
Based on this expression, one can verify that
\begin{eqnarray*}
&&\cov(\varphi_{1i}-F_gS_g^{-1}g_i)\\
&=&S_{\varphi_1}-F_gS_g^{-1}F_g^\textsf{T}\\
&=&S_{\varphi_1}-F_\gamma S_B^{-1}F_\gamma^\textsf{T}-(F_\gamma
S_B^{-1}H_{\gamma}^\textsf{T}-F_h)(S_h-H_{\gamma}S_B^{-1}H_{\gamma}^\textsf{T})^{-1}(F_\gamma
S_B^{-1}H_{\gamma}^\textsf{T}-F_h)^\textsf{T}.
\end{eqnarray*}

\noindent{\bf Proof of Theorem \ref{elwbound}} If  $\pi(x,\gamma^*)=\omega(x)$, recall that $\lambda^*=0$. According to the proof of Theorem \ref{influence}, we only need to show that
\begin{eqnarray}\label{v310}
F_gS_g^{-1}g_i=\frac{\delta_i-\pi(x_i,\gamma^*)}{\pi(x_i,\gamma^*)}u(x_i,\beta^*,\alpha^*).
\end{eqnarray}
Define
\begin{eqnarray*}
&&\xi_1(t_i,\beta^*,\gamma^*,\alpha^*)=\frac{\delta_i-\pi(x_i,\gamma^*)}{\pi(x_i,\gamma^*)}u(x_i,\beta^*,\alpha^*),\\ &&\xi_2(t_i,\beta^*,\gamma^*,\alpha^*)=\left(\frac{\delta_i-\pi(x_i,\gamma^*)}{\pi(x_i,\gamma^*)}a^\textsf{T}(x_i,\beta^*,\alpha^*),U_{1}^\textsf{T}(t_i,\gamma^*)\right)^\textsf{T}.
\end{eqnarray*}
Utilizing the identity (\ref{identity}), we have
\begin{eqnarray*}
F_gS_g^{-1}g_i&=&F_gS_g^{-1}\left(\begin{array}{c}
h_i\\
U_{1i}
\end{array}\right)\\
&=&F_{\gamma}(-S_B)^{-1}U_{1i}-F_gS_g^{-1}
\left(\begin{array}{c}
H_\gamma(-S_B)^{-1}U_{1i}-h_i\\
0
\end{array}\right).
\end{eqnarray*}
One can verify that $F_{\gamma}=E\left(\frac{\partial \varphi_{1}}{\partial  \gamma^\textsf{T}} \right)=E(\frac{\partial \xi_1}{\partial\gamma^\textsf{T}})=-E(\varphi_{1}U_1^\textsf{T})=-E(\xi_1U_1^\textsf{T})$ and $F_g=E(\varphi_{1}g^\textsf{T})=(E\{(1-\pi)u^{\otimes2}/\pi\},E(\varphi_{1}\xi_2^\textsf{T}))$. Moreover,
\begin{eqnarray*}
&&S_g^{-1}=\left(\begin{array}{cc}
E\{(1-\pi)u^{\otimes2}/\pi\}&E(\varphi_{1}\xi_2^\textsf{T})\\
E(\xi_2\varphi_{1}^\textsf{T})&E(\xi_2^{\otimes2})
\end{array}\right)^{-1}=\left(\begin{array}{cc}
D_{11}&D_{12}\\
D_{21}&D_{22}
\end{array}\right)^{-1}\\
&=&\left(\begin{array}{cc}
D_{11.2}^{-1}&-D_{11.2}^{-1}D_{12}D_{22}^{-1}\\
-D_{22}^{-1}D_{21}D_{11.2}^{-1}&D_{22}^{-1}+D_{22}^{-1}D_{21}D_{11.2}^{-1}D_{12}D_{22}^{-1}
\end{array}\right),
\end{eqnarray*}
where $D_{11.2}=D_{11}-D_{12}D_{22}^{-1}D_{21}$. Summarizing the above results, (\ref{v310}) is proved by noting that
\begin{eqnarray*}
&&F_gS_g^{-1}
\left(\begin{array}{c}
H_\gamma(-S_B)^{-1}U_{1i}-h_i\\
0
\end{array}\right)\\
&=&(-I_r,0)\left(\begin{array}{c}
H_\gamma(-S_B)^{-1}U_{1i}-h(t_i,\beta^*,\gamma^*,\alpha^*)\\
0
\end{array}\right)\\
&=&E\left(\frac{\partial \xi_1}{\partial\gamma^\textsf{T}}\right)(-S_B)^{-1}U_{1i}-\xi_1(t_i,\beta^*,\gamma^*,\alpha^*)\\
&=&F_\gamma(-S_B)^{-1}U_{1i}-\xi_1(t_i,\beta^*,\gamma^*,\alpha^*).
\end{eqnarray*}
Moreover, from (\ref{ifc}) and (\ref{v310}), we conclude that $V_{31}(\vartheta^*)=E(\varphi_2^{\otimes2})=E\{\pi^{-1}(g-u)^{\otimes2}\}+E(u^{\otimes2})$.

\noindent{\bf Proof of Theorem \ref{el}}
From the standard EL theory (Qin, 1994), we have
\begin{eqnarray}\label{ifgmm}
n^{1/2}\left(\begin{array}{c}
\hat{\beta}_{EL}-\beta^*\\
\hat{\gamma}_{EL}-\gamma^*\\
\end{array}\right)=S_{*}^{-1}S_{21}S_{11}^{-1}n^{-1/2}\sum_{i=1}^n\psi_{*}(t_i,\beta^*,\gamma^*)+o_p(1)\drow N(0,S_{*}^{-1}),
\end{eqnarray}
where $\psi_{*}(t_i,\beta^*,\gamma^*)=(\varphi_{1i}^\textsf{T},g_i^\textsf{T})^\textsf{T}$
$S_{11}=E\{\psi_{*}(t_i,\beta^*,\gamma^*)\psi_{*}^\textsf{T}(t_i,\beta^*,\gamma^*)\}$, $S_{21}=E\left\{\frac{\partial \psi_{*}^\textsf{T}(t_i,\beta^*,\gamma^*)}{\partial(\beta^\textsf{T},\gamma^\textsf{T})^\textsf{T}}\right\}$
and $S_{*}=S_{21}S_{11}^{-1}S_{21}^\textsf{T}=\cov(S_{21}S_{11}^{-1}\psi_{*}(t_i,\beta^*,\gamma^*))$. Repeated applications of the identity (\ref{identity}) yield
\begin{eqnarray*}
&&S_{21}S_{11}^{-1}\psi_{*}(t_i,\beta^*,\gamma^*)\\
&=&\left(\begin{array}{cc}
F_{\beta}^\textsf{T}&0\\
F_{\gamma}^\textsf{T}&G_{\gamma}^\textsf{T}
\end{array}\right)\left(\begin{array}{cc}
S_{\varphi_1}&F_g\\
F_{g}^\textsf{T}&S_{g}
\end{array}\right)^{-1}\left(\begin{array}{c}
\varphi_{1i}\\
g_i
\end{array}\right)\\
&=&\left(\begin{array}{cc}
F_{\beta}^\textsf{T}&0\\
F_{\gamma}^\textsf{T}&G_{\gamma}^\textsf{T}
\end{array}\right)\left(\begin{array}{cc}
S_{11.2}^{-1}&-S_{11.2}^{-1}F_gS_{g}^{-1}\\
-S_{g}^{-1}F_g^\textsf{T}S_{11.2}^{-1}&S_{g}^{-1}+S_{g}^{-1}F_g^\textsf{T}S_{11.2}^{-1}F_gS_{g}^{-1}
\end{array}\right)\left(\begin{array}{c}
\varphi_{1i}\\
g_i
\end{array}\right)\\
&=&\left(\begin{array}{cc}
F_{\beta}^\textsf{T}S_{11.2}^{-1}&-F_{\beta}^\textsf{T}S_{11.2}^{-1}F_gS_{g}^{-1}\\
0&G_{\gamma}^\textsf{T}S_{g}^{-1}
\end{array}\right)\left(\begin{array}{c}
\varphi_{1i}\\
g_i
\end{array}\right)\\
&=&\left(\begin{array}{c}
F_{\beta}^\textsf{T}S_{11.2}^{-1}(\varphi_{1i}-F_gS_{g}^{-1}g_i)\\
G_{\gamma}^\textsf{T}S_{g}^{-1}g_i
\end{array}\right)=\left(\begin{array}{c}
F_{\beta}^\textsf{T}S_{11.2}^{-1}(\varphi_{1i}-F_gS_{g}^{-1}g_i)\\
-U_{1i}
\end{array}\right),
\end{eqnarray*}
where $S_{11.2}=S_{\varphi_1}-F_gS_{g}^{-1}F_g^\textsf{T}=\cov(\varphi_{1i}-F_gS_{g}^{-1}g_i)$. Based on this fact, it follows that
\begin{eqnarray*}
S_{*}&=&\cov\left(\begin{array}{c}
F_{\beta}^\textsf{T}S_{11.2}^{-1}(\varphi_{1i}-F_gS_{g}^{-1}g_i)\\
-U_{1i}
\end{array}\right)=\left(\begin{array}{cc}
F_{\beta}^\textsf{T}S_{11.2}^{-1}F_{\beta}&0\\
0&S_{B}
\end{array}\right).
\end{eqnarray*}
Therefore,
\begin{eqnarray*}
S_{*}^{-1}&=&\left(\begin{array}{cc}
F_{\beta}^\textsf{T}S_{11.2}^{-1}F_{\beta}&0\\
0&S_{B}
\end{array}\right)^{-1}=
\left(\begin{array}{cc}
\Sigma_{1}&0\\
0&S_B^{-1}\\
\end{array}\right).
\end{eqnarray*}

\newpage
\begin{table}\begin{center}
\tiny
Table 1: Empirical bias  and RMSE (in parentheses) of $\beta$ in Model  1 with $n=200$ and different
missingness rates based on 1000 simulations.\\
\label{tab:1}       
\begin{tabular}{c ccc ccc cc}
 \hline
$(\tau_0,\tau_1,\tau_2,\tau_3)$   &Estimator &$k=1, \ \ \beta=3$ & $k=2, \ \ \beta=6$& $k=4, \ \ \beta=12$  \\
\hline
&&\multicolumn{3}{c}{True model $E(y|x)=2+3x_1^k+x_2^2$}\\
&&\multicolumn{3}{c}{Working model $E(y|x)=\alpha_0+\alpha_1x_1^2+\alpha_2x_2^2$}\\
$(1,0,0,0)$ &$\hat{\beta}_{ALL}$       &-0.0050 (0.0650)   &-0.0131 (0.1069)  &0.0062 (4.5214)             \\
                   &$\hat{\beta}_{CCA}$&0.0044 (0.0869)   &-0.0051 (0.1478)  &0.0192 (6.5074)             \\
$Miss\approx 0.27$ &$\hat{\beta}_{HT}$    &0.0020 (0.0701)   &0.0053 (0.1501)  &0.0713 (6.9084)             \\
                   &$\hat{\beta}_{RRZ}$&-0.0025 (0.0670)   &-0.0118 (0.1087)  &-0.0602 (4.9950)             \\
                   &$\hat{\beta}_{TQ}$ &-0.0260 (0.0685)   &-0.1067 (0.1202)  &-0.6057 (4.7712)             \\
                   &$\hat{\beta}_{QSZ}$&-0.0025 (0.0668)   &-0.0120 (0.1090)  &-0.0594 (4.9591)             \\
                  &$\hat{\beta}_{EDR}$ &-0.0020 (0.0683)   &-0.0116 (0.1087)  &-0.0231 (5.0291)             \\\\

$(1,0.5,0.5,0)$ &$\hat{\beta}_{ALL}$   &-0.0090 (0.0647)    &-0.0102 (0.1134)   &0.1431 (4.3659)             \\
                   &$\hat{\beta}_{CCA}$&0.3637 (0.2183)    &-0.1081 (0.1594)   &-0.3826 (5.2952)             \\
$Miss\approx 0.29$ &$\hat{\beta}_{HT}$ &-0.0031 (0.0781)    &0.0010 (0.1839)   &0.0347 (8.2344)             \\
                   &$\hat{\beta}_{RRZ}$&0.0041 (0.0767)    &-0.0162 (0.1140)   &-0.0188 (4.8512)             \\
                   &$\hat{\beta}_{TQ}$ &-0.0149 (0.0684)    &-0.1651 (0.1424)   &-0.8558 (4.3709)             \\
                   &$\hat{\beta}_{QSZ}$&-0.0026 (0.0728)    &-0.0163 (0.1138)   &-0.1343 (4.4735)             \\
                   &$\hat{\beta}_{EDR}$&0.0044 (0.0715)    &-0.0160 (0.1132)   &-0.0923 (4.5028)             \\\\

$(0.5,-0.5,0.5,0)$ &$\hat{\beta}_{ALL}$&0.0057 (0.0619)   &0.0001 (0.0993)  &0.0636 (4.2577)             \\
                   &$\hat{\beta}_{CCA}$&-0.5369 (0.3859)   &-0.0528 (0.1688)  &-0.1738 (6.6881)             \\
$Miss\approx 0.39$ &$\hat{\beta}_{HT}$ &0.0028 (0.0873)   &0.0191 (0.2597)  &0.0529 (10.4208)             \\
                   &$\hat{\beta}_{RRZ}$&-0.0107 (0.0919)   &0.0029 (0.1062)  &-0.0570 (5.1083)             \\
                   &$\hat{\beta}_{TQ}$ &-0.0929 (0.0766)   &-0.2096 (0.1516)  &-1.1846 (5.3331)             \\
                   &$\hat{\beta}_{QSZ}$&-0.0154 (0.0848)   &0.0034 (0.1080)  &-0.2363 (4.7173)             \\
                  &$\hat{\beta}_{EDR}$ &-0.0096 (0.0727)   &0.0035 (0.1063)  &-0.1865 (4.6021)             \\\\

$(0.5,0.5,1,0)$ &$\hat{\beta}_{ALL}$   &0.0108 (0.0672)    &0.0072 (0.1028)   &0.0630 (4.1344)             \\
                   &$\hat{\beta}_{CCA}$&0.4404 (0.2857)    &-0.0660 (0.1777)   &-0.1980 (6.6527)             \\
$Miss\approx 0.40 $ &$\hat{\beta}_{HT}$&-0.0045 (0.1212)    &-0.0154 (0.3132)   &-0.1884 (10.5059)             \\
                   &$\hat{\beta}_{RRZ}$&0.0238 (0.1236)    &0.0044 (0.1127)   &-0.1512 (4.7251)             \\
                   &$\hat{\beta}_{TQ}$ &-0.0390 (0.0735)    &-0.2629 (0.1826)   &-1.2918 (5.4722)             \\
                   &$\hat{\beta}_{QSZ}$&0.0155 (0.1409)    &0.0044 (0.1184)   &-0.3518 (4.4931)             \\
                   &$\hat{\beta}_{EDR}$&0.0300 (0.0947)    &0.0046 (0.1130)   &-0.2269 (4.3853)             \\

\hline
&&\multicolumn{3}{c}{True model $E(y|x)=2+3x_1^k+x_2^2$}\\
&&\multicolumn{3}{c}{Working model $E(y|x)=\alpha_0+\alpha_1x_1^k+\alpha_2x_2^2$}\\

$(0,0.5,1,1)$& $\hat{\beta}_{ALL}$     &0.0033 (0.0568)    &-0.0158 (0.1049)   &0.0044 (4.2293)             \\
                   &$\hat{\beta}_{CCA}$&0.5071 (0.3670)    &-0.1449 (0.2254)   &-0.4970 (7.6826)             \\
$Miss\approx 0.51 $ &$\hat{\beta}_{HT}$&-0.4076 (0.3262)    &0.5512 (1.4764)   &2.2205 (46.1843)             \\
                   &$\hat{\beta}_{RRZ}$&0.0007 (0.0812)    &-0.0114 (0.1280)   &0.0073 (4.2283)             \\
                   &$\hat{\beta}_{TQ}$ &-0.1100 (0.0796)    &-0.4158 (0.2889)   &-1.9422 (7.2001)             \\
                   &$\hat{\beta}_{QSZ}$&0.0029 (0.0675)    &-0.0136 (0.1198)   &0.0005 (4.2440)             \\
                   &$\hat{\beta}_{EDR}$&0.0008 (0.0627)    &-0.0166 (0.1171)   &0.0012 (4.2247)             \\\\

$(-1,0.5,1,1)$&$\hat{\beta}_{ALL}$    &0.0052 (0.0577)    &0.0044 (0.1125)   &0.0176 (4.2433)             \\
                  &$\hat{\beta}_{CCA}$&1.1173 (1.4338)    &0.2605 (0.4499)   &0.7233 (17.0767)             \\
$Miss\approx 0.69 $&$\hat{\beta}_{HT}$&-0.8450 (1.6744)    &1.7268 (10.9352)   &8.0815 (439.0630)             \\
                  &$\hat{\beta}_{RRZ}$&0.0227 (0.5527)    &0.0222 (0.4273)   &-0.0007 (4.4422)             \\
                  &$\hat{\beta}_{TQ}$ &-0.1286 (0.1037)    &-0.5903 (0.5139)   &-2.6869 (11.3583)             \\
                  &$\hat{\beta}_{QSZ}$&0.0073 (0.0825)    &0.0014 (0.1701)   &0.0208 (4.3328)             \\
                  &$\hat{\beta}_{EDR}$&0.0061 (0.0697)    &0.0054 (0.1479)   &0.0202 (4.2950)             \\

 \hline
\end{tabular}\end{center}
\end{table}

\newpage
\begin{table}\begin{center}
\tiny
Table 2: Empirical bias  and RMSE (in parentheses) of $\beta$ in Model  2 with $n=200$ and different
missingness rates based on 1000 simulations.\\
\label{tab:1}       
\begin{tabular}{cccc ccc }
 \hline
$(\tau_0,\tau_1,\tau_2)$ &Estimator &$k=1, \ \ \beta=2$ & $k=2, \ \ \beta=5$& $k=4, \ \ \beta=11$   \\
\hline
&&\multicolumn{3}{c}{True model $E(y|x)=2+3x^k$}\\
&&\multicolumn{3}{c}{Working model $E(y|x)=\alpha_0+\alpha_1x^2$}\\
$(1,0,0)$&$\hat{\beta}_{ALL}$         &-0.0002 (0.2181)    &0.0041 (0.3086)    &0.0690 (2.1748)  \\
                  &$\hat{\beta}_{CCA}$&-0.0043 (0.2598)    &0.0098 (0.3572)    &0.1323 (2.5779)  \\
$Miss\approx  0.27 $&$\hat{\beta}_{HT}$&-0.0004 (0.2192)    &-0.0024 (0.3542)    &0.0806 (2.5833)  \\
                  &$\hat{\beta}_{RRZ}$&-0.0008 (0.2200)    &-0.0115 (0.3080)    &-0.0488 (2.2223)  \\
                  &$\hat{\beta}_{TQ}$ &-0.0004 (0.2219)    &-0.0221 (0.3103)    &-0.2379 (2.1414)  \\
                  &$\hat{\beta}_{EDR}$&-0.0058 (0.2310)    &-0.0118 (0.3079)    &-0.0757 (2.2027)  \\\\

$(1,1,0)$&$\hat{\beta}_{ALL}$         &0.0061  (0.2189)    &0.0001 (0.3021)    &-0.0063 (2.1281)            \\
                  &$\hat{\beta}_{CCA}$&0.7653  (0.8033)    &-0.2283 (0.4109)    &-1.1650 (2.5382)            \\
$Miss\approx 0.30 $ &$\hat{\beta}_{HT}$&0.0398 (0.2500)    &-0.0791 (0.4570)    &-0.5038 (4.0154)            \\
                  &$\hat{\beta}_{RRZ}$&0.0690  (0.3173)    &-0.0256 (0.3009)    &-0.2466 (2.5654)            \\
                  &$\hat{\beta}_{TQ}$ &0.0419  (0.2271)    &-0.1070 (0.3198)    &-1.0523 (2.2353)            \\
                  &$\hat{\beta}_{EDR}$&0.0803  (0.2570)    &-0.0326 (0.3038)    &-0.5341 (2.1069)            \\\\

$(0.5,0.5,0)$&$\hat{\beta}_{ALL}$     &0.0164  (0.2184)    &-0.0241 (0.3102)    &-0.0663 (2.0871)            \\
                  &$\hat{\beta}_{CCA}$ &0.5536  (0.6156)    &-0.0700 (0.3956)    &-0.2758 (2.7491)            \\
$Miss\approx  0.38 $&$\hat{\beta}_{HT}$&0.0243  (0.2305)    &-0.0563 (0.4295)    &-0.3574 (2.8091)            \\
                  &$\hat{\beta}_{RRZ}$&0.0367  (0.2562)    &-0.0418 (0.3164)    &-0.3021 (2.2214)            \\
                  &$\hat{\beta}_{TQ}$ &0.0345  (0.2224)    &-0.0904 (0.3298)    &-0.7640 (2.1469)            \\
                  &$\hat{\beta}_{EDR}$&0.0479  (0.2360)    &-0.0439 (0.3179)    &-0.3999 (2.1027)            \\\\

$(0,0.5,0)$&$\hat{\beta}_{ALL}$       &-0.0036  (0.2092)    &-0.0001 (0.3028)    &-0.0280 (2.0204)            \\
                  &$\hat{\beta}_{CCA}$&0.7101  (0.7693)    &-0.0173 (0.4341)    &-0.0506 (2.9823)            \\
$Miss\approx  0.50 $&$\hat{\beta}_{HT}$&0.0102  (0.2368)    &-0.0841 (0.4818)    &-0.5596 (3.3812)            \\
                  &$\hat{\beta}_{RRZ}$&0.0306  (0.2901)    &-0.0189 (0.3044)    &-0.2757 (2.4513)            \\
                  &$\hat{\beta}_{TQ}$ &0.0242  (0.2196)    &-0.1093 (0.3285)    &-1.0708 (2.2621)            \\
                  &$\hat{\beta}_{EDR}$&0.0428  (0.2313)    &-0.0240 (0.3064)    &-0.5017 (2.1676)            \\

\hline
&&\multicolumn{3}{c}{True model $E(y|x)=2+3x^k$}\\
&&\multicolumn{3}{c}{Working model $E(y|x)=\alpha_0+\alpha_1x^k$}\\

$(0,1,1)$&$\hat{\beta}_{ALL}$         &0.0003  (0.2184)    &0.0152 (0.3219)    &-0.0705 (2.0708)            \\
                  &$\hat{\beta}_{CCA}$&0.4792  (0.5685)    &0.8412 (0.9580)    &3.6362 (4.7853)            \\
$Miss\approx  0.35 $&$\hat{\beta}_{HT}$&-0.2004  (0.2960)    &1.1282 (1.2449)    &5.5586 (7.2810)            \\
                  &$\hat{\beta}_{RRZ}$&0.0023  (0.2213)    &0.0063 (0.3234)    &-0.1155 (2.0496)            \\
                  &$\hat{\beta}_{TQ}$ &0.0024  (0.2201)    &0.0159 (0.3246)    &-0.0654 (2.0652)            \\
                  &$\hat{\beta}_{EDR}$&0.0025  (0.2184)    &0.0128 (0.3268)    &-0.1288 (2.0556)            \\\\

$(-1,1,1)$&$\hat{\beta}_{ALL}$        &-0.0019  (0.2175)    &0.0070 (0.3008)    &0.0866 (2.1650)            \\
                  &$\hat{\beta}_{CCA}$&0.9260  (1.0059)    &1.6656 (1.7628)    &8.0185 (9.2383)            \\
$Miss\approx  0.55  $&$\hat{\beta}_{HT}$&-0.7496 (0.8459)    &2.4618 (2.6430)    &13.8496 (17.3162)            \\
                  &$\hat{\beta}_{RRZ}$&-0.0107  (0.2330)    &0.0018 (0.3155)    &0.0394 (2.1534)            \\
                  &$\hat{\beta}_{TQ}$ &0.0016  (0.2296)    &0.0102 (0.3104)    &0.1122 (2.1743)            \\
                  &$\hat{\beta}_{EDR}$&0.0005  (0.2258)    &-0.0005 (0.3057)    &0.0327 (2.1326)            \\

 \hline
\end{tabular}\end{center}
\end{table}

\newpage
\begin{table}\begin{center}
\tiny
Table 3: Empirical bias  and RMSE (in parentheses) of $\beta$ in Model 3 with $n=200$ and  different
missingness rates based on 1000 simulations.
\\
\label{tab:1}       
\begin{tabular}{cccc ccc }
 \hline
$(\tau_0,\tau_1,\tau_2)$  &Estimator &$\beta_1=1$ & $\beta_2=1$& $\beta_3=1$   \\
\hline
$(1,0,0,0)$        &$\hat{\beta}_{ALL}$  &-0.0020   (0.1132)    &-0.0019 (0.0733)  &0.0021  (0.0512)    \\
                     &$\hat{\beta}_{CCA}$&-0.0007   (0.1311)    &-0.0039 (0.0865)  &0.0023  (0.0598)    \\
$Miss\approx  0.27$   &$\hat{\beta}_{HT}$&-0.0035   (0.1283)    &-0.0041 (0.0846)  &0.0044  (0.0601)    \\
                     &$\hat{\beta}_{RRZ}$&-0.0062   (0.1243)    &-0.0019 (0.0812)  &0.0059  (0.0627)    \\
                     &$\hat{\beta}_{TQ}$ &0.0003   (0.1273)    &-0.0047 (0.0864)  &0.0089  (0.0638)    \\
                     &$\hat{\beta}_{QZL}$&0.0064   (0.1257)    &-0.0094 (0.0846)  &0.0032  (0.0589)    \\
                     &$\hat{\beta}_{EDR}$&-0.0154   (0.1264)    &0.0014 (0.0827)  &0.0221  (0.0688)    \\\\

$(-3,2,2,-1)$        &$\hat{\beta}_{ALL}$&0.0052   (0.1139)    &-0.0010 (0.0733)  &-0.0027  (0.0502)    \\
                     &$\hat{\beta}_{CCA}$&0.5295   (0.5503)    &-0.2672 (0.2961)  &-0.0816  (0.0997)    \\
$Miss\approx  0.33$   &$\hat{\beta}_{HT}$&0.0763   (0.2609)    &-0.0653 (0.1838)  &-0.0113  (0.0734)    \\
                     &$\hat{\beta}_{RRZ}$&0.0339   (0.1840)    &-0.0535 (0.1449)  &0.0159  (0.0780)    \\
                     &$\hat{\beta}_{TQ}$ &0.1236   (0.1927)    &-0.1257 (0.1822)  &-0.0383  (0.0776)    \\
                     &$\hat{\beta}_{QZL}$&0.0190   (0.1819)    &-0.0606 (0.1457)  &0.0106  (0.0700)    \\
                     &$\hat{\beta}_{EDR}$&0.0329   (0.1458)    &-0.0640 (0.1337)  &0.0094  (0.0643)    \\\\

$(-2,2,2,-1)$        &$\hat{\beta}_{ALL}$&0.0021   (0.1159)    &0.0006 (0.0730)  &-0.0013  (0.0499)    \\
                     &$\hat{\beta}_{CCA}$&0.3674   (0.3917)    &-0.2002 (0.2286)  &-0.0554  (0.0771)    \\
$Miss\approx  0.22$   &$\hat{\beta}_{HT}$&0.0518   (0.2140)    &-0.0459 (0.1501)  &-0.0064  (0.0626)    \\
                     &$\hat{\beta}_{RRZ}$&0.0133   (0.1586)    &-0.0315 (0.1223)  &0.0186  (0.0644)    \\
                     &$\hat{\beta}_{TQ}$ &0.0925   (0.1703)    &-0.0918 (0.1505)  &-0.0269  (0.0701)    \\
                     &$\hat{\beta}_{QZL}$&0.0225   (0.1613)    &-0.0423 (0.1233)  &0.0086  (0.0612)    \\
                     &$\hat{\beta}_{EDR}$&0.0150   (0.1345)    &-0.0381 (0.1122)  &0.0127  (0.0597)    \\\\

$(-4,2,2,-1)$        &$\hat{\beta}_{ALL}$&0.0023   (0.1167)    &-0.0010 (0.0727)  &-0.0012  (0.0532)    \\
                     &$\hat{\beta}_{CCA}$&0.7090   (0.7301)    &-0.3244 (0.3583)  &-0.1052  (0.1243)    \\
$Miss\approx  0.48$   &$\hat{\beta}_{HT}$&0.1482   (0.3073)    &-0.1041 (0.2135)  &-0.02151  (0.0841)    \\
                     &$\hat{\beta}_{RRZ}$&0.0805   (0.2256)    &-0.0854 (0.1757)  &0.0181  (0.0846)    \\
                     &$\hat{\beta}_{TQ}$ &0.1386   (0.2153)    &-0.1513 (0.2080)  &-0.0424  (0.0905)    \\
                     &$\hat{\beta}_{QZL}$&0.0467   (0.1991)    &-0.0786 (0.1774)  &0.0155  (0.0792)    \\
                     &$\hat{\beta}_{EDR}$&0.0510   (0.1749)    &-0.0918 (0.1687)  &0.0159  (0.0751)    \\\\

$(-2,2,2,-2)$        &$\hat{\beta}_{ALL}$&-0.0014   (0.1076)    &0.0008 (0.0720)  &0.0005  (0.0504)    \\
                     &$\hat{\beta}_{CCA}$&0.5152   (0.5467)    &-0.5623 (0.6253)  &-0.0648  (0.0953)    \\
$Miss\approx  0.50$   &$\hat{\beta}_{HT}$&0.1036   (0.2927)    &-0.2034 (0.4285)  &-0.0054  (0.0878)    \\
                     &$\hat{\beta}_{RRZ}$&0.0530   (0.2145)    &-0.2258 (0.3395)  &0.0781  (0.1275)    \\
                     &$\hat{\beta}_{TQ}$ &0.2540   (0.2971)    &-0.3334 (0.3742)  &-0.0749  (0.1202)    \\
                     &$\hat{\beta}_{QZL}$&0.0541   (0.2294)    &-0.2303 (0.3517)  &0.0422  (0.1185)    \\
                     &$\hat{\beta}_{EDR}$&0.0737   (0.2484)    &-0.2131 (0.3170)  &0.0403  (0.1115)    \\
 \hline
\end{tabular}\end{center}
\end{table}

\newpage
\begin{table}\begin{center}
\tiny
Table 6: Empirical variance (EV), the mean of the variance estimators (MV) and the coverage probabilities of 95\%  confidence intervals (CP) of $\beta$ in Model 3 with $n=200$ and different missingness rates based on 1000 simulations.
\\
\label{tab:1}       
\begin{tabular}{cccc ccc ccc ccc }
 \hline
$(\tau_0,\tau_1,\tau_2)$ &Estimator &\multicolumn{3}{c}{$\beta_1=1$} & \multicolumn{3}{c}{$\beta_2=1$}& \multicolumn{3}{c}{$\beta_3=1$}   \\
\cmidrule(l){3-5}\cmidrule(l){6-8}\cmidrule(l){9-11}
& &  EV& MV & CP & EV& MV & CP& EV& MV & CP\\
\hline
$(1,0,0,0)$&$\hat{\beta}_{HT}$        &0.0164&0.0159&0.941 &0.0071&0.0063&0.918  &0.0036&0.0034&0.914    \\
$Miss\approx 0.27$&$\hat{\beta}_{RRZ}$&0.0154&0.0149&0.940 &0.0065&0.0062&0.915  &0.0039&0.0037&0.922    \\
                  &$\hat{\beta}_{EDR}$&0.0157&0.0222&0.955 &0.0068&0.0119&0.951  &0.0042&0.0064&0.940    \\\\

$(-3,2,2,-1)$&$\hat{\beta}_{HT}$      &0.0623&0.0313&0.793 &0.0295&0.0165&0.820  &0.0052&0.0037&0.897    \\
$Miss\approx 0.33$&$\hat{\beta}_{RRZ}$&0.0327&0.0411&0.901 &0.0181&0.0172&0.886  &0.0058&0.0055&0.916    \\
                  &$\hat{\beta}_{EDR}$&0.0202&0.0169&0.907 &0.0137&0.0106&0.853  &0.0040&0.0034&0.909    \\\\

$(-2,2,2,-1)$&$\hat{\beta}_{HT}$      &0.0431&0.0228&0.837 &0.0204&0.0121&0.833 &0.0038&0.0031&0.908 \\
$Miss\approx 0.22$&$\hat{\beta}_{RRZ}$&0.0250&0.0244&0.926 &0.0139&0.0121&0.899 &0.0038&0.0037&0.922 \\
                  &$\hat{\beta}_{EDR}$&0.0178&0.0146&0.916 &0.0111&0.0082&0.874 &0.0034&0.0031&0.914    \\\\

$(-4,2,2,-1)$&$\hat{\beta}_{HT}$      &0.0725&0.0403&0.707 &0.0348&0.0210&0.747 &0.0066&0.0043&0.868 \\
$Miss\approx 0.48$&$\hat{\beta}_{RRZ}$&0.0444&0.0506&0.855 &0.0236&0.0207&0.849 &0.0068&0.0063&0.927 \\
                  &$\hat{\beta}_{EDR}$&0.0280&0.0222&0.870 &0.0200&0.0149&0.813 &0.0053&0.0039&0.887    \\\\

$(-2,2,2,-2)$&$\hat{\beta}_{HT}$      &0.0750&0.0366&0.776 &0.1423&0.0551&0.638 &0.0076&0.0052&0.882 \\
$Miss\approx 0.50$&$\hat{\beta}_{RRZ}$&0.0432&0.0408&0.911 &0.0643&0.0525&0.762 &0.0101&0.0073&0.801 \\
                  &$\hat{\beta}_{EDR}$&0.0563&0.0433&0.853 &0.0551&0.0425&0.751  &0.0108&0.0066&0.816    \\

 \hline
\end{tabular}\end{center}
\end{table}

\newpage
\begin{table}\begin{center}
\tiny
Table 4: Empirical variance (EV), the mean of the variance estimators (MV) and the coverage probabilities of 95\%  confidence intervals (CP) of $\beta$ in Model  1 with $n=200$ and   different missingness rates based on 1000 simulations.\\
\label{tab:1}       
\begin{tabular}{ccc ccc ccc ccc}
 \hline
\multicolumn{1}{c}{$(\tau_0,\tau_1,\tau_2,\tau_3)$}&Estimator &\multicolumn{3}{c}{$k=1, \ \ \beta=3$}&\multicolumn{3}{c}{$k=2, \ \ \beta=6$}&\multicolumn{3}{c}{$k=4, \ \ \beta=12$}\\
\cmidrule(l){3-5}\cmidrule(l){6-8}\cmidrule(l){9-11}
& & EV& MV & CP & EV& MV & CP& EV& MV & CP\\
\hline

&&\multicolumn{9}{c}{True model $E(y|x)=2+3x_1^k+x_2^2$}\\
&&\multicolumn{9}{c}{Working model $E(y|x)=\alpha_0+\alpha_1x_1^2+\alpha_2x_2^2$}\\
$(1,0,0,0)$&$\hat{\beta}_{HT}$         &0.0702 &0.0649 &0.950 &0.1483 &0.1450 &0.938 &6.9102 &6.6814 &0.868    \\
$Miss\approx 0.27 $&$\hat{\beta}_{RRZ}$&0.0670 &0.0613 &0.941 &0.1079 &0.1070 &0.937 &4.9963 &4.7491 &0.878     \\
                &$\hat{\beta}_{EDR}$   &0.0683 &0.0935 &0.950 &0.1080 &0.1086 &0.937 &5.0336 &5.7043 &0.884       \\\\

$(1,0.5,0.5,0)$&$\hat{\beta}_{HT}$ &0.0709&0.0696&0.955&0.1876 &0.1675 &0.926     &8.4236 &7.2606 &0.857        \\
$Miss\approx 0.29$&$\hat{\beta}_{RRZ}$&0.0706&0.0687&0.943 &0.1207 &0.1081 &0.933  &4.9554 &4.3339 &0.862           \\
               &$\hat{\beta}_{EDR}$&0.0669&0.0642&0.941 &0.1209 &0.1076 &0.934     &4.4785 &3.9464 &0.856       \\\\

$(0.5,-0.5,0.5,0)$&$\hat{\beta}_{HT}$ &0.0852 &0.0803 &0.948 &0.2229&0.2037 &0.925 &10.2224 &8.9758 &0.837            \\
$Miss\approx 0.39$&$\hat{\beta}_{RRZ}$&0.0906 &0.0806 &0.945 &0.1120&0.1090 &0.948 &5.6184 &4.7022 &0.859           \\
                  &$\hat{\beta}_{EDR}$&0.0697 &0.0646 &0.946 &0.1122&0.1082 &0.949 &4.8118 &4.0836 &0.849           \\\\

$(0.5,0.5,1,0)$&$\hat{\beta}_{HT}$ &0.1284&0.1042&0.940&0.3686&0.2753&0.913    &13.2791 &10.6664 &0.830         \\
$Miss\approx 0.40 $&$\hat{\beta}_{RRZ}$&0.1055&0.1004&0.940&0.1088&0.1102&0.956  &5.5986 &4.7274 &0.868           \\
               &$\hat{\beta}_{EDR}$&0.0852&0.0859&0.937&0.1100&0.1082&0.951     &4.8260 &4.0707 &0.857        \\

\hline
&&\multicolumn{9}{c}{True model $E(y|x)=2+3x_1^k+x_2^2$}\\
&&\multicolumn{9}{c}{Working model $E(y|x)=\alpha_0+\alpha_1x_1^k+\alpha_2x_2^2$}\\

$(0,0.5,1,1)$&$\hat{\beta}_{HT}$       &0.1630&0.1419&0.769  &1.1402   &0.8671  &0.929   &41.2947&33.0218&0.866        \\
$Miss\approx 0.51 $&$\hat{\beta}_{RRZ}$&0.0850&0.0818&0.948  &0.1298   &0.1241  &0.942   &4.6293&4.3177&0.883         \\
                  &$\hat{\beta}_{EDR}$ &0.0650&0.0640&0.949  &0.1211   &0.1139  &0.941   &4.5905&4.3045&0.881        \\\\

$(-1,0.5,1,1)$&$\hat{\beta}_{HT}$     &0.9347&0.4978&0.621 &7.1428&3.3792&0.834  &538.6855&175.7416&0.789        \\
$Miss\approx 0.69 $&$\hat{\beta}_{RRZ}$&0.3333&0.3133&0.951 &0.4272&0.4455&0.943  &4.8705&4.7237&0.896        \\
                  &$\hat{\beta}_{EDR}$&0.0732&0.0681&0.949 &0.1482&0.1182&0.926  &4.6804&4.4822&0.896        \\

 \hline
\end{tabular}\end{center}
\end{table}

\newpage
\begin{table}\begin{center}
\tiny
Table 5: Empirical variance (EV), the mean of the variance estimators (MV) and the coverage probabilities of 95\%  confidence intervals (CP) of $\beta$ in Model 2 with $n=200$ and different missingness rates based on 1000 simulations.\\
\label{tab:1}       
\begin{tabular}{ccc ccc ccc ccc}
 \hline
\multicolumn{1}{c}{$(\tau_0,\tau_1,\tau_2)$}&Estimator &\multicolumn{3}{c}{$k=1, \ \ \beta=2$}&\multicolumn{3}{c}{$k=2, \ \ \beta=5$}&\multicolumn{3}{c}{$k=4, \ \ \beta=11$}\\
\cmidrule(l){3-5}\cmidrule(l){6-8}\cmidrule(l){9-11}
 & & EV& MV & CP & EV& MV & CP& EV& MV & CP\\
\hline
&&\multicolumn{9}{c}{True model $E(y|x)=2+3x^k$}\\
&&\multicolumn{9}{c}{Working model $E(y|x)=\alpha_0+\alpha_1x^2$}\\
$(1,0,0)$&$\hat{\beta}_{HT}$           &0.0481&0.0475&0.952 &0.1256&0.1254&0.935 &6.6739&6.1183&0.878 \\
$Miss\approx 0.27 $&$\hat{\beta}_{RRZ}$&0.0484&0.0478&0.953 &0.0948&0.0921&0.935 &4.9414&4.4330&0.876 \\
                &$\hat{\beta}_{EDR}$   &0.0533&0.0665&0.957 &0.0948&0.0922&0.936 &4.8513&4.6152&0.873 \\\\

$(1,1,0)$&$\hat{\beta}_{HT}$         &0.0597 &0.0566 &0.953 &0.2070 &0.1709 &0.892 &10.7677 &8.1392 &0.758           \\
$Miss\approx  0.30 $&$\hat{\beta}_{RRZ}$&0.1080 &0.0871 &0.924 &0.0968 &0.0930 &0.940 &5.3100 &4.2934 &0.843           \\
                 &$\hat{\beta}_{EDR}$&0.0588 &0.0486 &0.914 &0.0973 &0.0923 &0.936 &4.0106 &3.4968 &0.840            \\\\

$(0.5,0.5,0)$&$\hat{\beta}_{HT}$       &0.0523 &0.0513 &0.951 &0.1710 &0.1615 &0.914 &7.4481 & 7.0028 &0.818      \\
$Miss\approx  0.38 $&$\hat{\beta}_{RRZ}$&0.0627 &0.0596 &0.948 &0.0964 &0.0925 &0.941 &4.8610 &4.2273 &0.841         \\
                  &$\hat{\beta}_{EDR}$&0.0527 &0.0514 &0.946 &0.0969 &0.0923 &0.937 &4.2390 &4.3848 &0.859       \\\\

$(0,0.5,0)$&$\hat{\beta}_{HT}$      &0.0611 &0.0555 &0.933 &0.2341 &0.2074 &0.887 &10.9931 &8.6810 &0.781           \\
$Miss\approx  0.50 $&$\hat{\beta}_{RRZ}$&0.0893 &0.0747 &0.926 &0.0965 &0.0942 &0.937 &5.4194 &4.2620 &0.824       \\
                  &$\hat{\beta}_{EDR}$ &0.0564 &0.0495 &0.930 &0.0984 &0.0937 &0.936 &4.2609 &3.8816 &0.828        \\
\hline
&&\multicolumn{9}{c}{True model $E(y|x)=2+3x^k$}\\
&&\multicolumn{9}{c}{Working model $E(y|x)=\alpha_0+\alpha_1x^k$}\\
$(0,1,1)$&    $\hat{\beta}_{HT}$       &0.0505&0.0474&0.856&0.2442&0.2372&0.355&30.5626&18.9311&0.736        \\
$Miss\approx  0.35 $&$\hat{\beta}_{RRZ}$&0.0535&0.0490&0.934&0.0910&0.0938&0.934&4.5103&4.3444&0.892        \\
                  &$\hat{\beta}_{EDR}$ &0.0522&0.0476&0.934&0.0906&0.0934&0.930&4.4904&4.3390&0.895        \\\\

$(-1,1,1)$&   $\hat{\beta}_{HT}$       &0.1479&0.1445&0.434&1.0188&0.8292&0.105&98.0244&77.1886&0.443        \\
$Miss\approx  0.55  $&$\hat{\beta}_{RRZ}$&0.0542&0.0541&0.948&0.1025&0.0969&0.929&4.2466&4.2079&0.900        \\
                  &$\hat{\beta}_{EDR}$ &0.0506&0.0493&0.956&0.0976&0.0928&0.935&4.2046&4.2275&0.901        \\

 \hline
\end{tabular}\end{center}
\end{table}

\begin{table}\begin{center}

Table 7:  Point estimates, bootstrap standard errors (in parentheses in the first row) and the asymptotic variance formula-based standard errors (in parentheses in the second row).
\\
\label{tab:1}       
\begin{tabular}{cccc ccc }
 \hline
Estimator &$\mu_1$ & $\mu_0$& $\Delta$   \\
\hline
$\hat{\beta}_{CCA}$      &6349.14    (600.71)    &4554.80  (333.40)  &1794.34  (663.08)    \\
$\hat{\beta}_{QSZ}$&6262.68    (602.95)    &4527.26  (336.58)  &1735.42  (664.98)    \\
$\hat{\beta}_{TQ}$&6216.76 (592.38)    &4394.82  (322.74)  &1821.94  (679.37)    \\
$\hat{\beta}_{HT}$ &6210.97    (595.32)    &4540.08  (337.56)  &1670.88  (656.88)    \\
                   &~~~~~~~~~~ (571.24)    &~~~~~~~~~~ (344.27)  &~~~~~~~~~~ (665.96)    \\
$\hat{\beta}_{RRZ}$&6263.55    (597.08)    &4523.00  (337.97)  &1740.55  (661.88)    \\
                   &~~~~~~~~~~ (575.99)    &~~~~~~~~~~ (347.82)  &~~~~~~~~~~ (671.61)    \\
$\hat{\beta}_{EDR}$&6262.65    (609.73)    &4547.66  (339.06)  &1714.99  (669.14)    \\
                   &~~~~~~~~~~ (588.46)    &~~~~~~~~~~ (344.35)  &~~~~~~~~~~ (684.75)    \\
 \hline
\end{tabular}\end{center}
\end{table}

\end{document}